\documentclass[english]{extarticle}
\usepackage[T1]{fontenc}
\usepackage[latin9]{inputenc}
\usepackage{geometry}
\usepackage{float}
\usepackage{amsmath}
\usepackage{amsthm}
\usepackage{amssymb}
\usepackage{graphicx}
\usepackage{esint}

\makeatletter
\numberwithin{equation}{section}
\numberwithin{figure}{section}
\newcommand{\lyxaddress}[1]{
	\par {\raggedright #1
	\vspace{1.4em}
	\noindent\par}
}

\numberwithin{equation}{section}

\makeatother

\usepackage{babel}
\begin{document}
\title{TCSA and the finite volume boundary state}
\author{Zoltan Bajnok, Tamas Lajos Tompa}
\maketitle

\lyxaddress{\begin{center}
\emph{Wigner Research Centre for Physics}\\
\emph{Konkoly-Thege Miklós u. 29-33, 1121 Budapest , Hungary}\\
\par\end{center}}
\begin{abstract}
We develop a new way to calculate the overlap of a boundary state
with a finite volume bulk state in the truncated conformal space approach.
We check this method in the thermally perturbed Ising model analytically,
while in the scaling Lee-Yang model numerically by comparing our results
to excited state g-functions, which we obtained by the analytical
continuation method. We also give a simple argument for the structure
of the asymptotic overlap between the finite volume boundary state
and a periodic multiparticle state, which includes the ratio of Gaudin
type determinants. 
\end{abstract}

\section{Introduction}

Recently there have been growing interest in integrable finite volume
boundary states and significant progress has been made in the calculation
of their overlaps with periodic multiparticle states. These research
are fueled by quantum quench problems in statistical physics as well
as correlation function calculations in the AdS/CFT duality.

In the AdS/CFT correspondence there are many places where integrable
overlaps appeared recently. At first a codimension one defect is introduced
in the gauge theory, which allows one-point functions of local gauge
invariant operators to be different from zero. These operators correspond
to finite volume multiparticle states on the worldsheet and their
one-point functions in the presence of the defect can be interpreted
as overlaps with a --hopefully integrable-- boundary state. There
were many perturbative calculations for different subsectors in the
D3-D5 \cite{deLeeuw:2015hxa,Buhl-Mortensen:2015gfd,deLeeuw:2018mkd}
and D3-D7 \cite{deLeeuw:2016ofj,deLeeuw:2019ebw} setup and for a
few loop orders in the coupling \cite{Buhl-Mortensen:2016pxs,Buhl-Mortensen:2017ind}
together with new developments also in bootstrapping the boundary
state and calculating asymptotic overlaps \cite{Linardopoulos:2020jck,Kristjansen:2020mhn,Gombor:2020kgu,Gombor:2020auk,Komatsu:2020sup}.
Recently, a new class of three-point functions were investigated involving
a local gauge invariant single trace operator and two determinant
operators dual to maximal giant gravitons and it was shown to be an
overlap between a finite volume multiparticle state and a finite volume
integrable boundary state \cite{Jiang:2019xdz,Jiang:2019zig}. 

In a statistical physical quench problem there is a sudden change
in one of the parameters of the theory. As a result the ground-state
of the pre-quenched theory is no longer an eigenstate of the post-quenched
one, thus initiates a non-trivial time evolution, which can be integrable
\cite{Caux:2013ra}. This evolution is relevant to understand if thermalization
happens and can be calculated once the overlaps between the (integrable)
initial state and multiparticle  states are known. The exact overlap
results in integrable spin chains \cite{Kozlowski:2012fv,Brockmann_2014_1,Pozsgay:2018ixm}
indicated that non-vanishing overlaps appear for parity symmetric
Bethe states only. This was connected to the integrability of the
boundary/initial state in \cite{Piroli:2017sei,Piroli:2018ksf,Piroli:2018don}.
There were further progress to calculate time evolutions after a quench
from the overlaps in \cite{Horvath:2015rya,Horvath:2017wzf,Horvath:2018gat,Rakovszky:2016ugs,Hodsagi:2019rcs}. 

In both applications above the exact knowledge of the overlaps between
a finite volume boundary state and periodic multiparticle states is
required. Not much is known about these overlaps in quantum field
theories for generic excited states. In \cite{Kormos:2010ae} the
first few large volume overlaps were extracted from the investigation
of the one-point function. The aim of our paper is twofold. On the
one hand we would like to extend this results and describe the finite
volume boundary state in terms of the infinite volume characteristics
of the theory such as reflection and scattering matrices for any finite
size. On the other hand we would like to test these formulas against
numerical data obtained from the truncated conformal space approach
(TCSA). Although most of the analysis is restricted to the simplest
integrable interactive theory, namely to the scaling Lee-Yang model
the conceptual ideas and generic form of the results can be easily
generalized. 

There were already two approaches to calculate excited state overlaps
(called g-functions there) from TCSA. The first was concerned with
massless flows \cite{Takacs:2011aa}, while the second extracted the
g-functions from the evaluation of the partition functions obtained
from the finite strip TCSA spectrum \cite{Dorey:1997yg,Dorey:1999cj}.
Since these approaches seem to be very cumbersome we decided to develop
a novel and more direct way of computing the overlaps in the TCSA
framework. We test this new method against the large volume scattering
description. 

Following an old idea, which goes back at least to Lüscher \cite{Luscher:1985dn,Luscher:1986pf},
it is possible to express finite volume quantities in terms of their
infinite volume counterparts. This program was successfully completed
for the finite size spectrum, it is under development for finite volume
form factors and it can be also carried out for the finite volume
boundary state as we will demonstrate in this paper for the scaling
Lee-Yang model.

Depending on the nature of the finite size corrections we can distinguish
three domains: The leading corrections are polynomial in the inverse
of the volume, $O(L^{-1})$, and originate from imposing periodic
boundary condition. Characteristic quantities, such as energies and
form factors take their infinite volume form, but periodicity implies
momentum quantization, which makes the spectrum discrete and changes
the natural normalization of states. Subleading corrections are exponentially
small in the volume and their leading term, $O(e^{-mL})$, contains
a pair of virtual particles which travel around the periodic world
\cite{Bajnok:2008bm}. For small enough volumes all virtual processes
have to be summed up, which is nicely realized by the Thermodynamic
Bethe Ansatz equation for the ground-state \cite{Zamolodchikov:1989cf}.
Exact finite volume energies for excited states can be obtained by
analytical continuation \cite{Dorey:1996re}. Here assumptions are
made for the analytical behavior for the pseudoenergies and the equations
obtained are conjectures which have to be tested.

The calculation of the overlap between the finite volume boundary
state and the finite volume ground-state called the g-function has
a long history. The g-function was first introduced in conformal field
theories and characterized the ground-state boundary entropy \cite{Affleck:1991tk}.
Later it was shown to decrease along renormalization group flows similarly
to the central charge of CFT \cite{Friedan:2003yc}. The first attempt
to calculate the g-function in massive integrable theories dealt with
the evaluation of the partition function with two boundaries in two
alternative ways related to each other by exchanging the role of space
and time \cite{LeClair:1995uf}. By evaluating the path integral in
the saddle point approximation the boundary condition dependent part
of the g-function was correctly calculated. However, the g-function
appears as the O(1) piece in the free energy which includes contributions
from the quadratic fluctuations as well as from the non-trivial measure
in constructing the functional integral \cite{Woynarovich:2004gc}.
The first result accounting for the full result was obtained in diagonally
scattering theories. It was conjectured based on the cluster expansion
of the partition function and on attempts to take into account the
O(1) contributions \cite{Dorey:2004xk}. Later, the measure in the
path integral was correctly calculated providing the honest derivation
of the ground-state g-function \cite{Pozsgay:2010tv}. Alternatively,
the complete summation of the cluster expansion leads to the same
results \cite{Kostov:2018dmi,Kostov:2019sgu}. The authors also tried
to extend this approach to excited states' g-functions, however the
incorporation of the asymptotic overlaps seemed problematic, although
recently improved. This was revealed by an extension of the results
for non-diagonal scattering theories and for excited states by analytical
continuation in the AdS/CFT setting \cite{Jiang:2019xdz,Jiang:2019zig}.
The aim of our paper is to elaborate further on the calculation of
boundary states by analytical continuation and also in the TCSA framework. 

The paper is organized as follows: In the next section we recall the
infinite volume characterization of integrable boundaries both in
space and in time. We then present the polynomial corrections for
the boundary state. This will be done after reviewing the same corrections
for the spectrum and for form factors. In Section 3 we recall the
g-function in diagonally scattering one species models and perform
an analytical continuation to obtain the excited states' g-functions
together with g-functions in the presence of boundary boundstates.
In Section 4 we present our novel proposal for the calculation of
excited state g-function in the TCSA framework. We test this proposal
in Section 5 for the thermally perturbed Ising model by exact analytical
calculations. Section 6 deals with the scaling Lee-Yang model where
we compare the novel TCSA results to the numerical evaluation of the
TBA g-functions, whose implementation is spelled out in Appendix B.
We finally conclude in Section 7 and provide some outlook. 

\section{The asymptotic boundary state}

We first review the two ways how an integrable boundary can be placed
in a infinite volume by restricting the 1+1 dimensional space-time
into the half plane. We then determine how the infinite volume quantities
and momentum quantization can be used to calculate the polynomial
corrections to the finite volume boundary state. 

\subsection{Infinite volume boundary state}

An integrable boundary can be placed either in space or in time \cite{Ghoshal:1993tm}. 

\subsubsection{Boundary in space, reflections}

If the boundary is in space (space is restricted to the negative half-line)
boundary conditions can be characterized by reflection factors. The
reflection matrix $R_{n\to m}$ is the amplitude of a process in which
an asymptotic initial $n$-particle state consisting of $n$ separated
right moving particles reflects into an asymptotic final $m$-particle
state consisting reflected left moving particles. In integrable theories
there is no particle creation and the multiparticle reflection factorizes
into the product of individual reflections and pairwise scatterings
\begin{equation}
R_{n\to n}(\theta_{1},\dots,\theta_{n})=\prod_{i<j}S(\theta_{i}-\theta_{j})\prod_{i=1}^{n}R(\theta_{i})
\end{equation}
where $\theta$ is the rapidity, which parameterizes the relativistic
dispersion relation as $(E(\theta),p(\theta))=m(\cosh\theta,\sinh\theta)$.
The integrable scattering matrix $S(\theta_{1}-\theta_{2})$ which
depends only on the rapidity differences satisfies unitarity and crossing
symmetry
\begin{equation}
S(\theta)S(-\theta)=1\quad;\qquad S(i\pi-\theta)=S(\theta)
\end{equation}
 The reflection factor also satisfies unitarity and boundary crossing
symmetry 
\begin{equation}
R(\theta)R(-\theta)=1\quad;\qquad R(i\pi-\theta)=S(2\theta)R(\theta)\label{eq:Rbootstrap}
\end{equation}
 The origin of the boundary crossing equation seems a bit obscure,
but will be clear from the other channel point of view. Once the scattering
matrix is specified solutions of the reflection equations characterize
integrable boundary conditions in the bootstrap setting. We will analyze
two models in this paper, the thermally perturbed Ising model and
the scaling Lee-Yang model.

The thermally perturbed Ising model is related to the theory of a
free massive fermion and has scattering matrix $S=-1$. The two solutions
of the (\ref{eq:Rbootstrap}) bootstrap equations which correspond
to fixed/free boundary conditions have the reflection factors 
\begin{equation}
R_{\pm}(\theta)=\left(\pm\frac{1}{2}\right)_{\theta}\quad;\qquad(x)_{\theta}=\frac{\sinh(\frac{\theta}{2}+\frac{i\pi x}{2})}{\sinh(\frac{\theta}{2}-\frac{i\pi x}{2})}
\end{equation}

The scaling Lee-Yang model is the only relevant perturbation of the
conformal Lee-Yang model and has the scattering matrix \cite{Cardy:1989fw}
\begin{equation}
S(\theta)=\frac{\sinh\theta+i\sin\frac{\pi}{3}}{\sinh\theta-i\sin\frac{\pi}{3}}
\end{equation}
We will be interested in two reflection factors, which are related
to perturbations where only the bulk is perturbed \cite{Dorey:1997yg}.
There are two conformal boundary conditions in the Lee-Yang model
labeled by $\mathbb{I}$ and $\varphi$. The bulk only perturbation
of the identity boundary condition is described by the reflection
factor 
\begin{equation}
R_{\mathbb{I}}(\theta)=\left(\frac{1}{6}\right)_{\theta}\left(\frac{1}{2}\right)_{\theta}\left(-\frac{2}{3}\right)_{\theta}\label{eq:Rid}
\end{equation}
while the perturbation of $\varphi$ by 
\begin{equation}
R_{\varphi}(\theta)=R_{\mathbb{I}}(\theta)\left(\frac{b-1}{6}\right)_{\theta}\left(\frac{b+1}{6}\right)_{\theta}\left(\frac{5-b}{6}\right)_{\theta}\left(-\frac{5+b}{6}\right)_{\theta}\label{eq:Rphi}
\end{equation}
with $b=-1/2$. Note that $R_{\mathbb{I}}(\theta)$ can be formally
obtained by putting $b=0$. If $b$ is not $-1/2$ then there is also
a perturbation at the boundary, but we will not discuss this situation
in the paper. 

\subsubsection{Boundary in time, the boundary state}

If the boundary is placed in time it can be characterized as a boundary
state \cite{Ghoshal:1993tm}. This translation invariant state can
be expanded in the basis of the Hilbert space associated to the whole
line as 
\begin{equation}
\vert B\rangle=\sum_{n=0}^{\infty}\frac{1}{n!}\prod_{i}\int\frac{d\theta_{i}}{4\pi}K_{n}(\theta_{1},\dots,\theta_{n})\vert\theta_{1},\dots,\theta_{n},-\theta_{n},\dots,-\theta_{1}\rangle\label{eq:Bstate_infvol}
\end{equation}
Infinite volume states are normalized as $\langle\theta\vert\theta'\rangle=2\pi\delta(\theta-\theta')$
and integrations go for the whole line. Due to relativistic invariance
in the bulk one can relate the boundary state to the multiparticle
reflection amplitude as
\begin{equation}
K_{n}(\theta_{1},\dots,\theta_{n})=R_{n\to n}(\frac{i\pi}{2}-\theta_{1},\dots,\frac{i\pi}{2}-\theta_{n})
\end{equation}
Integrability guaranties that only pairs of particles with opposite
rapidities appear and the extra $1/2$ in the $1/(4\pi)$ measure
reflects this fact. Factorized scattering implies that $K_{n}$ can
be written in terms of the scattering matrix and the reflection factor.
Reordering the multiparticle terms shows that the boundary state exponentiates
\begin{equation}
\vert B\rangle=\exp\{\int\frac{d\theta}{4\pi}K(\theta)Z(-\theta)Z(\theta)\}\vert0\rangle
\end{equation}
where $Z(\theta)$ is the ZF operator, which creates a particle with
rapidity $\theta$ and form an exchange algebra 
\begin{equation}
Z(\theta_{1})Z(\theta_{2})=S(\theta_{1}-\theta_{2})Z(\theta_{2})Z(\theta_{1})
\end{equation}
while 
\begin{equation}
K(\theta)=K_{2}(\theta)=R(\frac{i\pi}{2}-\theta)
\end{equation}
Consistency of the boundary state implies the relation $K(\theta)=S(2\theta)K(-\theta)$
which is the origin of the boundary crossing unitarity relation. The
overlap of an infinite volume multiparticle state with the boundary
state is then 
\begin{equation}
\langle-\theta_{1},\theta_{1},\dots,-\theta_{n},\theta_{n}\vert B\rangle=\prod_{j=1}^{n}K(\theta_{i})
\end{equation}

Let us point out that there is a subtlety here. If the reflection
factor has a pole at $i\frac{\pi}{2}$, or equivalently if the $K$-
matrix has a pole at $\theta=0$ of the form 
\begin{equation}
K(\theta)=-\frac{ig^{2}}{2\theta}+\mathrm{regular}
\end{equation}
then the boundary state has a one particle contribution \cite{Ghoshal:1993tm,Dorey:1997yg,Dorey:1998kt,Bajnok:2006dn}:
\begin{equation}
\vert B\rangle=\left\{ 1+\frac{g}{2}Z(0)\right\} \exp\left\{ \fint\frac{d\theta}{4\pi}K(\theta)Z(-\theta)Z(\theta)\right\} \vert0\rangle
\end{equation}
with a principal value integration. This one-particle term has a drastic
effect on the finite size energy corrections on the strip \cite{Bajnok:2004tq}
and on boundary form factors \cite{Bajnok:2006ze}, and also shows
up in the overlaps with odd number of particles (containing a standing
particle):
\begin{equation}
\langle-\theta_{1},\theta_{1},\dots,-\theta_{n},\theta_{n},0\vert B\rangle=\frac{g}{2}\prod_{j=1}^{n}K(\theta_{i})\label{eq:Infinitevolumeoverlap}
\end{equation}

\subsection{Polynomial corrections to the finite volume boundary state}

In the following we give a simple argument for the form of the asymptotic
finite volume boundary state. We start by recalling the similar corrections
for the energy spectrum and also for form factors.

\subsubsection{Leading finite size corrections for the spectrum and for form factors}

As we already mentioned in the introduction we express all finite
volume quantities in terms of their infinite volume analogues. Putting
$N$ particles in a finite volume $L$ the momentum is quantized by
the Bethe-Yang (BY) equations
\begin{equation}
Q_{j}=p(\theta_{j})L-i\sum_{k:k\neq j}\log S(\theta_{j}-\theta_{k})=2\pi n_{j}
\end{equation}
Once the quantization numbers $\{n_{j}\}$ are specified the rapidities
$\{\theta_{j}\}$ can be calculated. For large volumes the free quantization
dominates $p(\theta_{i})=\frac{2\pi n_{i}}{L}$ and the scattering
interaction introduces polynomial corrections in $L^{-1}$. The energy
of the state containing all of these polynomial corrections is formally
the same as in infinite volume 
\begin{equation}
E(n_{1},\dots,n_{N})=\sum_{j}E(\theta_{j})+O(e^{-mL})
\end{equation}
but the momenta are quantized by the BY equations.

We now recall the leading finite size correction of the form factor
\cite{Pozsgay:2007gx,Pozsgay:2007kn}, which is the matrix element
of a local operator $\mathcal{O}$ between asymptotic states: 
\begin{equation}
F_{N}^{\mathcal{O}}(\theta_{1},\dots,\theta_{N})=\langle0\vert\mathcal{O}\vert\theta_{1},\dots,\theta_{N}\rangle
\end{equation}
This state is originally defined for strictly ordered rapidities,
$\theta_{i}>\theta_{i+1}$, but can be generalized via $\vert\theta_{1},\dots,\theta_{i},\theta_{i+1}\dots,\theta_{N}\rangle=S(\theta_{i}-\theta_{i+1})\vert\theta_{1},\dots,\theta_{i+1},\theta_{i},\dots,\theta_{N}\rangle$
for any orderings. The normalization is 
\begin{equation}
\langle\theta_{N}',\dots,\theta_{1}'\vert\theta{}_{1},\dots,\theta{}_{N}\rangle=(2\pi)^{N}\prod_{i=1}^{N}\delta(\theta_{i}-\theta'_{i})+\mbox{permutations}
\end{equation}
where permutations contains terms obtained by exhanging $\theta$s
all possible ways and by picking up the corresponding S-matrix factors. 

The finite volume state can be characterized either by the quantization
numbers $\{n_{i}\}$ or by the corresponding rapidities $\{\theta_{i}\}$.
We denote this state as $\vert n_{1},\dots,n_{N}\rangle_{L}\equiv\vert\theta_{1},\dots,\theta_{N}\rangle_{L}$,
which is typically symmetric for the exchanges of particles. Since
the finite volume spectrum is discrete the natural normalization is
\begin{equation}
_{L}\langle n_{1}',\dots,n_{N}'\vert n_{1},\dots,n_{N}\rangle_{L}=\prod_{i}\delta_{n_{i}'n_{i}}
\end{equation}
We emphasize that it is not the same as the infinite volume normalization,
which can be easily seen by relating them for large volumes. In doing
so we compare the resolution of the identity
\begin{equation}
\sum_{N=0}^{\infty}\sum_{\{n_{i}\}}\vert n_{1},\dots,n_{N}\rangle_{\!L}\,_{L}\!\langle n_{N},\dots,n_{1}\vert=\mathbb{I}=\sum_{N=0}^{\infty}\prod_{i}\int\frac{d\theta_{i}}{2\pi}\vert\theta_{1},\dots,\theta_{N}\rangle\langle\theta_{N},\dots,\theta_{1}\vert
\end{equation}
The two operators on the two sides of the equation act in different
Hilbert spaces, however for large volumes the rapidities become spaced
as dense as $\frac{2\pi}{L}$. The very dense rapidity space summation
for the $N$ particles states then can be replaced with integration
\begin{equation}
\sum_{\{n_{i}\}}\vert\{n_{i}\}\rangle_{L}\,_{L}\langle\{n_{i}\}\vert=\prod_{i}\int\frac{d\theta_{i}}{2\pi}\rho_{N}(\{\theta_{j}\})\vert\theta_{1},\dots,\theta_{N}\rangle_{L}\,\,_{L}\!\langle\theta_{N},\dots,\theta_{1}\vert
\end{equation}
where the density of states is the Jacobian of changing variables
from $\{n_{j}\}$ to $\{\theta_{j}\}$ via the BY equations:
\begin{equation}
\rho_{N}(\{\theta_{j}\})=\mbox{det}\left|\frac{\partial Q_{j}}{\partial\theta_{i}}\right|=\mbox{det}\left|\begin{array}{ccc}
p'_{1}L+\phi_{12}+\dots\phi_{1N} & \dots & -\phi_{1N}\\
\vdots & \ddots & \vdots\\
-\phi_{N1} & \dots & p'_{N}L+\phi_{N1}+\dots+\phi_{NN-1}
\end{array}\right|
\end{equation}
where $p'(\theta)=\frac{dp(\theta)}{d\theta}$ and $\phi_{jk}=\phi(\theta_{j}-\theta_{k})=-i\partial_{\theta_{j}}\log S(\theta_{j}-\theta_{k})$.
We can now compare the integrands for each particle numbers' integrations
leading to the relation between finite and infinite volume normalizations
\cite{Pozsgay:2007kn}
\begin{equation}
\vert\theta_{1},\dots,\theta_{N}\rangle_{L}=\frac{\vert\theta_{1},\dots,\theta_{N}\rangle}{\sqrt{\prod_{i<j}S(\theta_{i}-\theta_{j})\rho_{N}(\theta_{1},\dots,\theta_{N})}}+O(e^{-mL})
\end{equation}
Clearly the phase of the state cannot be fixed in this way, it follows
from demanding a symmetric finite volume state. We note that the relation
does not rely on large particle numbers, merely on large volumes,
and is valid even for one or two particle states. 

The finite volume form factor up to exponentially small corrections
is basically the infinite volume form factor modulo the change in
the normalization of states: 
\begin{equation}
\langle0\vert\mathcal{O}\vert\theta_{1},\dots,\theta_{N}\rangle_{L}=\frac{F_{N}^{\mathcal{O}}(\theta_{1},\dots,\theta_{N})}{\sqrt{\prod_{i<j}S(\theta_{i}-\theta_{j})\rho_{N}(\theta_{1},\dots,\theta_{N})}}+O(e^{-mL})
\end{equation}
as was obtained from comparing the infinite and finite volume 2-point
functions in \cite{Pozsgay:2007kn}. Let us use this normalization
change to calculate the finite volume boundary state.

\subsubsection{Leading finite size correction for the boundary state}

The finite volume boundary state can be expressed in the basis of
the finite volume Hilbert space with periodic boundary condition in
the form: 
\begin{equation}
\vert B\rangle_{L}=\vert B\rangle_{L}^{\mathrm{even}}+\vert B\rangle_{L}^{\mathrm{odd}}
\end{equation}
where we separated the contribution of the parity symmetric BY states
for containing even and odd number of particles. The odd term $\vert B\rangle_{L}^{\mathrm{odd}}$
is nonzero only if $\mathrm{res}_{\theta=0}K(\theta)\neq0$. We spell
out this term in Appendix \ref{sec:GenBstate}. In the following we
assume that this term is absent and write simply 
\begin{equation}
\vert B\rangle_{L}=\sum_{N}\sum_{\{n_{j}\}}K_{N}(n_{1},\dots,n_{N})_{L}\vert n_{1},\dots,n_{N},-n_{N},\dots,-n_{1}\rangle_{L}
\end{equation}
Similarly how we related the resolution of the identity for large
volumes we can also replace the summation for pairs of particles for
integrations of those pairs. In doing so we have to use the relation
\begin{equation}
Q_{j}^{-}=p(\theta_{j})L-i\sum_{k:k\neq j}\log S(\theta_{j}-\theta_{k})-i\sum_{k}\log S(\theta_{j}+\theta_{k})=2\pi n_{j}
\end{equation}
and change variable from $\{n_{j}\}$ to $\{\theta_{j}\}$. The corresponding
Jacobian is $\rho_{N}^{-}=\mbox{\ensuremath{\det\left|\frac{\partial Q_{j}^{-}}{\partial\theta_{i}}\right|}}$.
Additionally we also express the finite volume state in terms of the
infinite volume state and arrive at 
\begin{equation}
\vert B\rangle_{L}=\sum_{N}\prod_{i}\int\frac{d\theta_{i}}{4\pi}K_{N}(\theta_{1},\dots,\theta_{N})_{L}\frac{\rho_{N}^{-}}{\sqrt{\prod_{i}S(-2\theta_{i})\rho_{2N}}}\vert-\theta_{1},\theta_{1},-\theta_{2},\theta_{2},\dots,-\theta_{N},\theta_{N}\rangle
\end{equation}
Here we note that the correct way of calculating the norm of the state
is to move a bit away from the pairwise structure and take the appropriate
limit. This can be easily done by introducing one single extra particle
and sending its rapidity to infinity. What is important is that we
have to consider $\theta$ and $-\theta$ independent when calculating
the derivatives and substitute their pairwise structure into the general
$\rho_{2N}$. Actually $\rho_{2N}$ for pairwise states $\{\theta_{1},\dots,\theta_{N},-\theta_{1},\dots,-\theta_{N}\}$
factorizes as 
\begin{equation}
\rho_{2N}=\mbox{det}\left|\begin{array}{cc}
A & B\\
B & A
\end{array}\right|=\mbox{det}\left|\begin{array}{cc}
A-B & B\\
0 & A+B
\end{array}\right|=\rho_{N}^{-}\rho_{N}^{+}
\end{equation}
where $\rho_{N}^{\pm}$ are obtained from $\rho_{N}$ by replacing
$\phi_{ij}=\phi(\theta_{i}-\theta_{j})$ with $\phi_{ij}^{\pm}=\phi(\theta_{i}-\theta_{j})\pm\phi(\theta_{i}+\theta_{j})$.
By comparing this expression with the infinite volume boundary state
(\ref{eq:Bstate_infvol}) we arrive at the relation
\begin{equation}
K_{N}(\theta_{1},\dots,\theta_{N})_{L}=\frac{\sqrt{\rho_{2N}}}{\sqrt{\prod_{i}S(2\theta_{i})}\rho_{N}^{-}}\prod_{i}K_{2}(\theta_{i})+O(e^{-mL})=\prod_{i}\frac{K_{2}(\theta_{i})}{\sqrt{S(2\theta_{i})}}\sqrt{\frac{\rho_{N}^{+}}{\rho_{N}^{-}}}+O(e^{-mL})\label{eq:Kas}
\end{equation}
where we assumed that, similarly to the spectrum and form factors,
we have no other polynomial volume effects. The first few terms of
this expression was obtained from the analysis of the 1-point function
in \cite{Kormos:2010ae} and the structure was proven for simple spin
chains in \cite{Jiang:2020sdw}. Our argumentations could be extended
for integrable spin chains, if one could show that the boundary state
in the thermodynamic limit has the form (\ref{eq:Bstate_infvol}).
Note that the finite volume boundary state is a symmetric even function
of the rapidities. We will confirm this result by calculating the
exact finite volume expressions and taking the large volume limit. 

\section{Exact finite volume boundary state}

In this section we recall the derivation of the exact finite volume
g-functions. We then review how to  extend the result for excited
states by analytical continuation.

\begin{figure}
\begin{centering}
\includegraphics[width=4cm]{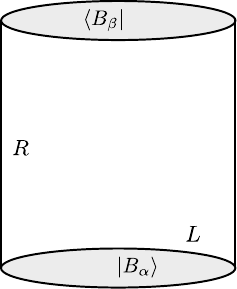}~~~~~~~~~~~~~~\includegraphics[height=4cm]{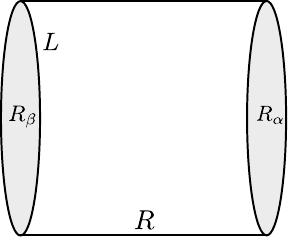}
\par\end{centering}
\caption{Two alternative ways to calculate the partition function depending
on the chosen time evolution.}

\label{cylinder}
\end{figure}

The finite volume groundstate overlap is called the g-function, which
can be obtained from the evaluation of the partition function on the
cylinder \cite{LeClair:1995uf}, see Figure \ref{cylinder}. Depending
on which time evolution we choose there are two representations of
the partition function
\begin{equation}
Z_{\alpha\beta}(L,R)=\,_{L}\langle B_{\beta}\vert e^{-H(L)R}\vert B_{\alpha}\rangle_{L}=\text{Tr}(e^{-H_{\alpha\beta}(R)L})
\end{equation}
 By evolving for a very large Euclidean time, $R\to\infty$, only
the periodic groundstate survives
\begin{equation}
\lim_{R\to\infty}Z_{\alpha\beta}(L,R)=\,_{L}\langle B_{\beta}\vert0\rangle\langle0\vert B_{\alpha}\rangle_{L}e^{-E_{0}(L)R}+\dots
\end{equation}
allowing for a direct extraction of the g-function $g_{\alpha}(L)=\langle0\vert B_{\alpha}\rangle_{L}$
as
\begin{equation}
\log(g_{\alpha}(L)g_{\beta}(L))=\lim_{R\to\infty}\left\{ R^{-1}\log Z_{\alpha\beta}(L,R)+E_{0}(L)\right\} 
\end{equation}
The idea is to calculate this quantity in the other channel, where
the volume is large and polynomial corrections are enough to be kept:
\begin{equation}
Z_{\alpha\beta}(L,R)=\text{Tr}(e^{-H_{\alpha\beta}(R)L})=\sum_{n}e^{-E_{\alpha\beta}^{n}(R)L}
\end{equation}
here $n$ labels multiparticle state on the strip. These polynomial
corrections come from momentum quantization with reflections on the
two ends and scatterings among the particles: 
\begin{equation}
2\pi n_{k}=2mR\sinh\theta_{k}-i\log R_{\alpha}(\theta_{k})-i\log R_{\beta}(\theta_{k})-i\sum_{j:j\neq k}\log\left\{ S(\theta_{k}-\theta_{j})S(\theta_{k}+\theta_{j})\right\} 
\end{equation}
The energy is 
\begin{equation}
E_{\alpha\beta}^{n}(R)=\sum_{k=1}^{n}m\cosh\theta_{k}+f_{\alpha}+f_{\beta}
\end{equation}
where $f_{\alpha}$ and $f_{\beta}$ are the boundary energies, which
are typically normalized to 0 in the TBA setting. Recently there were
developments in summing up directly the contributions of the multi-particle
states in the cluster expansion \cite{Kostov:2018dmi}. This can be
done by introducing non-linear integral equations for the diagrammatic
evaluation of the contributions. Alternatively, the traditional way
is to introduce densities of states and turn the sum for multiparticle
states into a functional integral for the densities \cite{LeClair:1995uf}.
Since the g-function is the O(1) term in the partition function special
care needs to be taken for constructing the measure of the integral
as well as for evaluating the quadratic fluctuations \cite{Woynarovich:2004gc,Dorey:2004xk,Pozsgay:2010tv}. 

\subsection{Groundstate g-function}

The outcome of these analysis is that the g-function can be written
as 
\begin{equation}
g_{\alpha}(L)g_{\beta}(L)=\frac{\det(1-K^{-})}{\det(1-K^{+})}\exp\left\{ \int\frac{d\theta}{4\pi}\left(\phi_{\alpha}(\theta)+\phi_{\beta}(\theta)-2\phi(2\theta)-2\pi\delta(\theta)\right)\log(1+e^{-\epsilon(\theta)})\right\} 
\end{equation}
where $\phi_{\alpha/\beta}(\theta)=-i\partial_{\theta}\log R_{\alpha/\beta}(\theta)$
and $\epsilon(\theta)$ is the solution of the TBA equation 
\begin{equation}
\epsilon(\theta)=mL\cosh\theta-\int\frac{d\theta'}{2\pi}\phi(\theta-\theta')\log(1+e^{-\epsilon(\theta')})
\end{equation}
for the groundstate energy
\begin{equation}
E_{0}^{\mathrm{TBA}}(L)=-m\int\frac{d\theta}{2\pi}\cosh\theta\log(1+e^{-\epsilon(\theta)})
\end{equation}
 on the circle of size $L$. The Fredholm operators $K^{\pm}$ are
defined by their integral kernels: 
\begin{equation}
K^{\pm}f(\theta)=\int_{0}^{\infty}\frac{d\theta'}{2\pi}\frac{\phi^{\pm}(\theta,\theta')}{1+e^{\epsilon(\theta)}}f(\theta')\quad;\qquad\phi^{\pm}(\theta,\theta')=\phi(\theta-\theta')\pm\phi(\theta+\theta')
\end{equation}
The numerator in the ratio of Fredholm determinants comes from the
fluctuations, while the denominator is related to the non-trivial
measure in the functional integral \cite{Pozsgay:2010tv,Woynarovich:2010wt}.

\subsection{Groundstate $g$-function with boundary boundstates}

This formula is valid if there are no boundary boundstates. In some
cases it can happen that the physical boundary condition characterized
by $R_{\alpha}(\theta)$ has a boundary boundstate say at $iu:$ 
\begin{equation}
R_{\alpha}(\theta)=i\frac{g_{\alpha}^{2}}{2(\theta-iu)}+\mathrm{regular}
\end{equation}
In this case the g-function takes the form \cite{Dorey:1999cj}
\begin{equation}
g_{\alpha}(L)=\sqrt{\frac{\det(1-K^{-})}{\det(1-K^{+})}}(1+e^{-\epsilon(iu)})\exp\left\{ \int\frac{d\theta}{4\pi}\left(\phi_{\alpha}(\theta)-\phi(2\theta)-\pi\delta(\theta)\right)\log(1+e^{-\epsilon(\theta)})\right\} 
\end{equation}
This term can be understood as an analytical continuation from a domain
where there are no boundary boundstates. During the continuation the
pole of $\phi_{\alpha}$ crosses the integration contour, whose contribution
has to be picked up. The g-function without this contribution corresponds
to the boundary boundstate. For more than one boundstates all of their
contributions $\prod_{j}(1+e^{-\epsilon(iu_{j})})$ have to be included. 

\subsection{Excited states g-function}

We can analytically continue the ground-state equations in order to
describe excited states. In doing so singularities $\bar{\theta}_{i}$,
related to the particles' rapidities, $\theta_{i}$, cross the integration
contour and introduce extra source terms \cite{Dorey:1996re}. For
the energy they contribute as 
\begin{equation}
E_{n}^{\mathrm{TBA}}(L)=im\sum_{j}\eta_{j}\sinh\bar{\theta}_{j}-m\int\frac{d\theta}{2\pi}\cosh\theta\,\log(1+e^{-\epsilon(\theta)})
\end{equation}
where $\epsilon$ satisfies the excited state TBA equation 
\begin{equation}
\epsilon(\theta)=mL\cosh\theta+\sum_{j}\eta_{j}\log S(\theta-\bar{\theta}_{j})-\int\frac{d\theta'}{2\pi}\phi(\theta-\theta')\log(1+e^{-\epsilon(\theta')})\label{eq:tbaex}
\end{equation}
and the parameters $\eta_{j}$ keep track of the location of the poles.
For poles on the upper half plane $\eta_{j}=1$, while for those on
the lower half plane $\eta_{j}=-1$. The position of the singularities
are related to the zeros of the logarithm of $1+e^{-\epsilon}$ :
\begin{equation}
\epsilon(\bar{\theta}_{k})=i\pi(2n_{k}+1)
\end{equation}
For free fermions and for the sinh-Gordon model $\bar{\theta}_{k}=\theta_{k}+i\frac{\pi}{2}$,
while for the scaling Lee-Yang model each particle is represented
by a complex conjugate pair of rapidities $\bar{\theta}_{k}=\theta_{k}+i\delta_{k}$
and $\bar{\theta}_{k+n}=\theta_{k}-i\delta_{k}$, where for large
volumes $\delta_{k}$ approaches $\pi/6$. The excited states g-functions
are the overlaps between the finite volume boundary state and periodic
finite volume multiparticle states
\begin{equation}
g_{\alpha}(\{n_{k}\},L)=\sqrt{\frac{\det(1-K_{\mathrm{ex}}^{-})}{\det(1-K_{\mathrm{ex}}^{+})}}\prod_{k}\frac{R_{\alpha}(\bar{\theta}_{k})}{\sqrt{S(2\bar{\theta}_{k})}}\exp\left\{ \int\frac{d\theta}{4\pi}\left(\phi_{\alpha}(\theta)-\phi(2\theta)-\pi\delta(\theta)\right)\log(1+e^{-\epsilon(\theta)})\right\} 
\end{equation}
where the Fredholm operators have a discrete and a continuous part
\cite{Bajnok:2019yik,Bajnok:2019mpp}: 
\begin{equation}
K_{\mathrm{ex}}^{\pm}.\left(\begin{array}{c}
f(\bar{\theta}_{j})\\
f(\theta')
\end{array}\right)=\left(\begin{array}{cc}
K_{k}^{\pm}(\bar{\theta}_{j}) & K^{\pm}(\bar{\theta}_{k},\theta')\\
K_{j}^{\pm}(\theta) & K^{\pm}(\theta,\theta')
\end{array}\right)\left(\begin{array}{c}
f(\bar{\theta}_{j})\\
f(\theta')
\end{array}\right)
\end{equation}
where 
\begin{equation}
K_{k}^{\pm}(\theta)=\frac{\phi^{\pm}(\bar{\theta}_{k},\theta)}{-i\partial_{\theta}\epsilon(\theta)\vert_{\bar{\theta}_{k}}}\quad;\qquad K^{\pm}(\theta,\theta')=\frac{\phi^{\pm}(\theta,\theta')}{1+e^{\epsilon(\theta')}}
\end{equation}
It is easy to see in the sinh-Gordon and free fermion cases that $R_{\alpha}(\theta+\frac{i\pi}{2})/\sqrt{S(i\pi+2\theta)}=K(\theta)/\sqrt{S(2\theta)}.$
It was shown in \cite{Jiang:2019xdz,Jiang:2019zig} that the large
volume limit of $\det(1-K_{\mathrm{ex}}^{\pm})$ goes to $\det\rho^{\mp}$
thus the exact result reproduces correctly the asymptotic form (\ref{eq:Kas}). 

For the free fermion the kernel vanishes, $\phi(\theta)=0$, and the
exact overlaps take the form 
\begin{equation}
g_{\alpha}(\{n_{k}\},L)=\prod_{k}iK_{\alpha}(\theta_{k})\exp\left\{ \int\frac{d\theta}{4\pi}(\phi_{\alpha}(\theta)-\pi\delta(\theta))\log(1+e^{-mL\cosh(\theta)})\right\} \label{eq:freeg}
\end{equation}

For the scaling Lee-Yang model we elaborate on the numerical evaluations
of the overlaps in Appendix \ref{sec:numTBA} as these will be compared
to the same quantities obtained from TCSA, which is the topic of the
next section.

\section{Overlaps in the TCSA framework}

In this section we propose a new way to ``measure'' the overlaps
between the finite volume boundary state and the finite volume multiparticle
states in the TCSA framework.

\subsection{TCSA for the energy}

The truncated conformal space approach is a very efficient variational
approximation, which was originally developed for relevant perturbations
of conformal field theories in order to calculate their finite size
energy spectrum on the circle \cite{Yurov:1989yu}. The Hamiltonian
of the perturbed system can be written as 
\begin{equation}
H=H_{0}+\lambda\int_{0}^{L}\Phi(x,t)dx\quad;\qquad H_{0}=\frac{2\pi}{L}(L_{0}+\bar{L}_{0}-\frac{c}{12})
\end{equation}
where $c$ is the central charge of the CFT, while the Virasoro generators
$L_{0}$ and $\bar{L}_{0}$ act diagonally on the Virasoro highest
weight representations $\mathcal{V}_{i}$, $\bar{\mathcal{V}}_{i}$
from which the Hilbert space is built up\footnote{Here we assume diagonal modular invariant partition functions.}
as 
\begin{equation}
\mathcal{H}=\sum_{i}\mathcal{V}_{i}\otimes\bar{\mathcal{V}}_{i}\label{eq:Hspace}
\end{equation}
By mapping the cylinder onto the plane, ($x+iy\to z=e^{-i\frac{2\pi}{L}(x+iy)}=re^{i\theta}$),
the Hamiltonian takes the form

\begin{equation}
H=\frac{2\pi}{L}\left[L_{0}+\bar{L}_{0}-\frac{c}{12}+\lambda\left(\frac{L}{2\pi}\right)^{2-2h}2\pi\int_{0}^{2\pi}\frac{d\theta}{2\pi}\,\Phi(e^{i\theta},e^{-i\theta})\right]
\end{equation}
where $h$ is the chiral weight of the spinless perturbation. The
$\theta$ integration implies momentum conservation. Typically, when
we calculate the matrix elements of $\Phi$, we choose the Virasoro
basis in each Verma module and ensure that singular vectors are eliminated.
This basis is not orthogonal and the inner product matrix, $G$, can
be calculated from the defining relations of the Virasoro algebra.
The idea of TCSA is to truncate the conformal Hilbert space at a given
energy $E_{cut}$, and to diagonalize numerically the Hamiltonian
\begin{equation}
\frac{H}{m}=\frac{2\pi}{mL}\left(L_{0}+\bar{L}_{0}-\frac{c}{12}+\Bigl(\frac{mL}{2\pi\kappa}\Bigr)^{2-2h}2\pi\,G^{-1}\Phi\right)\label{eq:HTCSA}
\end{equation}
at the given truncation, where 
\begin{equation}
m=\kappa\lambda^{\frac{1}{2-2h}}\label{eq:kappa}
\end{equation}
is a mass parameter, typically chosen to be the massgap. The eigenvalues
$E_{n}(L)$ of the Hamiltonian are the finite volume energy levels
of multiparticle states, while the eigenvectors $\vert n,\lambda\rangle_{L}$
are their representations on the conformal Hilbert space. These vectors
then can be used to calculate finite volume form factors\footnote{TCSA can be used to calculate boundary form factors as well \cite{Kormos:2007qx,Lencses:2011ab}.}
\cite{Pozsgay:2007kn} and, as we will show, also finite volume overlaps.

Alternatively one can also do perturbation theory with the Hamiltonian
\begin{equation}
H=H_{0}+\lambda L^{1-2h}V
\end{equation}
This is called conformal perturbation theory (CPT) and for the groundstate
energy it gives
\begin{equation}
E_{0}(L,\lambda)=-\frac{\pi c}{6L}+\lambda\,\langle\text{vac}\vert V(L)\vert\text{vac}\rangle+\dots=-\frac{\pi c}{6L}+L^{-1}\sum_{k=1}^{\infty}(\lambda L^{2-2h})^{k}b_{k}
\end{equation}
while for the groundstate eigenvector 
\begin{equation}
\vert\text{vac},\lambda\rangle_{L}=\vert\text{vac}\rangle+\lambda L^{2-2h}\sum_{n\neq\text{vac}}c_{n}\vert n\rangle+\dots\quad;\qquad c_{n}L=\frac{\langle n\vert V\vert\text{vac}\rangle}{\langle\text{vac}\vert H_{0}\vert\text{vac}\rangle-\langle n\vert H_{0}\vert n\rangle}
\end{equation}
where $\vert\text{vac}\rangle$ denotes the lowest energy state of
the conformal field theory. The perturbative expansion of the groundstate
energy indicates that it is normalized differently than the groundstate
energy in the TBA setting 
\begin{equation}
E_{0}(L,\lambda)=E_{0}^{\mathrm{TBA}}(L)+f_{\mathrm{bulk}}L\label{eq:fbulk}
\end{equation}
the difference being the vacuum energy density.

\subsection{Conformal boundary conditions}

In a conformal field theory we can also place a boundary in two different
ways \cite{Cardy:1989fw}. A space boundary is conformal, if there
is no energy flow through the boundary: $T_{xy}\vert_{x=0}=0$. In
holomorphic coordinates it reads as $T(z)\vert_{x=0}=\bar{T}(\bar{z})\vert_{x=0}$,
which implies that the symmetry is no longer the tensor product of
two Virasoro algebras, rather it is one single Virasoro algebra. Boundary
conditions can be labeled by representations of the chiral algebra,
such that on the strip with boundary conditions $\alpha$ and $\beta$
representations of the fusion product of $\alpha$ and $\beta$ appear
only in the decomposition of the Hilbert space
\begin{equation}
\mathcal{H}_{\alpha\beta}=\sum_{i}N_{\alpha\beta}^{i}\mathcal{V}_{i}
\end{equation}
where $N_{\alpha\beta}^{i}$ are the Verlinde fusion numbers which
can be calculated from the modular $\mathcal{S}$-matrix. 

If however the boundary is in time it is represented as an initial
or final state, $\vert B\rangle$, which can be expanded in the basis
of the periodic Hilbert space, obtained by mapping the boundary onto
the plane. Conformal invariance translates into the condition
\begin{equation}
(L_{n}-\bar{L}_{-n})\vert B\rangle=0
\end{equation}
These conformal states, called Ishibashi states, are unique for each
highest weight representation $\mathcal{V}_{i}\otimes\bar{\mathcal{V}_{i}}$
and can be constructed as follows: We choose the basis in each chiral
half by Virasoro descendants including only linearly independent combinations
\begin{equation}
\vert i\rangle=L_{-n_{1}}\dots L_{-n_{k}}\vert h\rangle\quad;\qquad\langle i\vert=\langle h\vert L_{n_{k}}\dots L_{n_{1}}
\end{equation}
Their inner product matrix follows from the Virasoro algebra and will
be denoted as 
\begin{equation}
\langle i\vert j\rangle=M_{ij}\quad;\qquad M^{ij}M_{jk}=\delta_{k}^{i}
\end{equation}
The Ishibashi state can be written simply in terms of the inverse
of this inner product matrix as
\begin{equation}
\vert h\rangle\!\rangle=M^{ik}\vert i\rangle\otimes\vert\bar{k}\rangle
\end{equation}
where $\vert\bar{k}\rangle$ is obtained from $\vert k\rangle$ by
replacing $L$ with $\bar{L}$. 

Space and time boundaries can be connected by conformal transformations.
Specific boundary conditions in space, labeled by representations,
do not correspond directly to Ishibashi states. They are related to
Cardy states $\vert B_{\alpha}\rangle$, which are linear combinations
of Ishibashi states
\begin{equation}
\vert B_{\alpha}\rangle=\sum_{i}g_{\alpha}^{i}\vert h_{i}\rangle\!\rangle
\end{equation}
The coefficients can be fixed by demanding that the cylinder partition
function, calculated in the two alternative time evolutions, are the
same. We normalize the Ishibashi states as 
\begin{equation}
\langle\!\langle h_{j}\vert e^{-H_{0}R}\vert h_{i}\rangle\!\rangle=\delta_{ij}\langle h_{i}\vert h_{i}\rangle\chi_{i}(q)
\end{equation}
Time evolution by the strip Hamiltonian $H_{\text{strip}}(R)=\frac{\pi}{R}(L_{0}-\frac{c}{24})$
on the Hilbert space $\mathcal{H}_{\alpha\beta}$ leads to Cardy's
condition
\begin{equation}
g_{\alpha}^{i}g_{\beta}^{i}\langle i\vert i\rangle=\sum_{j}N_{\alpha\beta}^{j}\mathcal{S}_{j}^{i}
\end{equation}
where $\mathcal{S}$ is the matrix which represents the modular transformation
on the characters. This equation can be solved, which provides the
physical, Cardy, boundary states \cite{Cardy:1989ir}.

\subsection{TCSA for boundary states and overlaps}

In the following we analyze theories which are perturbed in the bulk
but not at the boundaries. The perturbed theory is a massive scattering
theory, which in the integrable case can be characterized by the mass
of the particle and by its scattering matrix. If the boundary is in
space these particles reflect off it and the boundary condition is
characterized by the reflection matrix. If however the boundary is
in time then it is characterized by its overlap with the bulk multiparticle
states. In the case when there is no boundary perturbation it is natural
to assume that the boundary state does not change\footnote{If in the bulk-boundary OPEs of the bulk perturbing operator there
are singularities, one might need to renormalize the boundary state.} and we can calculate the overlap as the overlap between the conformal
boundary state, i.e. the Cardy state, and the normalized multiparticle
state, $\vert n,\lambda\rangle$, represented on the conformal Hilbert
space either by TCSA or by CPT. Thus we define the overlap in the
deformed theory for the vacuum as 
\begin{equation}
g_{\alpha}^{\mathrm{TCSA}}(L)=\langle\text{vac},\lambda\vert B_{\alpha}\rangle\label{eq:tcsag0}
\end{equation}
By comparing the small coupling/volume perturbative formulas \cite{Dorey:1999cj}
we can conclude that it is normalized differently than the $g$-function
in the TBA setting. They differ by a term linear in the volume, which
is proportional to the boundary energy $f_{\alpha}$: 
\begin{equation}
\log g_{\alpha}^{\mathrm{TCSA}}(L)=\log g_{\alpha}(L)-f_{\alpha}L\label{eq:logf}
\end{equation}
The excited states overlaps, which can be compared to the TBA results
are simply 
\begin{equation}
\log g_{\alpha}(n,L)=\log\langle n,\lambda\vert B_{\alpha}\rangle+f_{\alpha}L\label{eq:tcsagex}
\end{equation}
In the following we test this idea analytically in the thermally perturbed
Ising model and numerically in the scaling Lee-Yang model.

\section{Free massive fermion}

The critical Ising model can be described by the $c=\frac{1}{2}$
CFT which has three representations $h=0$, $h=\frac{1}{2}$ and $h=\frac{1}{16}$.
Its spinless perturbation with the $h=\frac{1}{2}$ field corresponds
to moving away from the critical temperature. This model is equivalent
to free massive fermions of mass $m$, which can be solved exactly
providing an ideal framework to test our proposal for the overlaps.

\subsection{Free massless fermion}

Let us first describe the CFT in terms of free massless (Euclidean)
fermions
\begin{equation}
i\psi(z)=\sum_{n}b_{n}z^{-n-\frac{1}{2}}\quad;\qquad\{b_{n},b_{m}\}=\delta_{n+m}
\end{equation}
They can be either periodic, Neveu-Schwartz sector, with half integer
modes or anti-periodic, Ramond sector, with integer modes. The Virasoro
algebra is represented in the NS sector as 
\begin{equation}
L_{n}=\sum_{j>\frac{n}{2}}(j-\frac{n}{2})b_{n-j}b_{j}\quad;\qquad j\in\mathbb{N}+\frac{1}{2}
\end{equation}
In the Ramond sector we simply take $j\in\mathbb{N}$ and for $j=0$
\begin{equation}
L_{0}=\sum_{j>0}jb_{-j}b_{j}+\frac{1}{4}b_{0}^{2}\quad;\qquad b_{0}^{2}=\frac{1}{2}
\end{equation}
There are analogous formulas for the other chiral fermion $\bar{\psi}(\bar{z})$
in terms of $\bar{b}_{n}$. The NS vacuum, $\vert0\rangle$, corresponds
to the identity representation of the Virasoro algebra, while the
$h=\frac{1}{2}$ representation to $b_{-\frac{1}{2}}\bar{b}_{-\frac{1}{2}}\vert0\rangle.$
The Ramond vacuum is nothing but the $h=\frac{1}{16}$ highest weight
state. In the following we work with periodic boundary condition and
take the free Hamiltonian 
\begin{equation}
H_{0}=\frac{2\pi}{L}(L_{0}+\bar{L}_{0}-\frac{c}{12})=\frac{2\pi}{L}(\sum_{n>0}nb_{-n}b_{n}+\sum_{n>0}n\bar{b}_{-n}\bar{b}_{n}-\frac{1}{24})
\end{equation}
in the NS sector. 

\subsection{Mass perturbation}

We then add the relevant spinless perturbation on the cylinder which
corresponds to the mass term
\begin{equation}
H=H_{0}+mV\quad;\qquad V=-\frac{i}{2\pi}\int_{0}^{L}\psi\bar{\psi}dx
\end{equation}
By mapping the perturbation onto the plane we get the perturbation
in terms of fermion modes: 
\begin{equation}
V=i\left(\frac{2\pi}{L}\right)^{2h-1}\int\frac{d\theta}{2\pi}\sum_{n,k}b_{n}\bar{b}_{k}e^{i\theta(n-k)}=i\sum_{n}b_{n}\bar{b}_{n}
\end{equation}
The $n^{th}$ mode Hilbert space with basis $\{\vert0\rangle,b_{-n}\vert0\rangle,\bar{b}_{-n}\vert0\rangle,b_{-n}\bar{b}_{-n}\vert0\rangle\}$
is left invariant by the perturbation, thus we can simply diagonalize
the corresponding 4 by 4 matrix. Since the momentum 
\begin{equation}
P_{0}=\frac{2\pi}{L}(L_{0}-\bar{L}_{0})=\frac{2\pi}{L}(\sum_{n>0}nb_{-n}b_{n}-\sum_{n>0}n\bar{b}_{-n}\bar{b}_{n})
\end{equation}
commutes with the perturbed Hamiltonian and it has eigenvalues $\{0,k_{n},-k_{n},0\}$
with $k_{n}=2n\pi/L$, only the zero momentum vectors $\vert0\rangle$
and $b_{-n}\bar{b}_{-n}\vert0\rangle$ are mixed. As a result the
eigenvalues are $\{k_{n}-\omega_{n},k_{n},k_{n},k_{n}+\omega_{n}\}$
with $\omega_{n}=\sqrt{m^{2}+k_{n}^{2}}$ and the energy differences
account correctly for the free relativistic spectrum of mass $m$.
Clearly the interacting vacuum is a complicated object in the conformal
Hilbert space as it is entangled between all the modes. The easiest
way do describe this vacuum is to introduce new creation-annihilation
operators with a given momentum eigenvalue $k_{n}$ for $n>0$: 
\begin{equation}
\beta_{n}=b_{n}+\gamma_{n}\bar{b}_{-n}\quad;\qquad\beta_{n}^{\dagger}=b_{-n}+\gamma_{n}^{\star}\bar{b}_{n}
\end{equation}
and similarly with momentum $-k_{n}$: 
\begin{equation}
\beta_{-n}=\alpha_{n}b_{-n}+\bar{b}_{n}\quad;\qquad\beta_{-n}^{\dagger}=\alpha_{n}^{\star}b_{n}+\bar{b}_{-n}
\end{equation}
These operators are not normalized properly, rather they are written
in the form, such that in the $m\to0$ limit both $\alpha_{n}\to0$
and $\gamma_{n}\to0$ and we get back the conformal result. By demanding
the free fermion form for the Hamiltonian 
\begin{equation}
H=\sum_{n>0}\text{\ensuremath{\left\{  \rho_{n}(\beta_{n}^{\dagger}\beta_{n}+\beta_{-n}^{\dagger}\beta_{-n})+\kappa_{n}\right\} } }
\end{equation}
together with the independence of the modes $\{\beta_{n},\beta_{-n}\}=0$
we can find the following solution 
\begin{equation}
\rho_{n}=\frac{m}{2}e^{\theta_{n}}\quad;\qquad\alpha_{n}=-ie^{-\theta_{n}}\quad;\quad\gamma_{n}=ie^{-\theta_{n}}\quad;\qquad\kappa_{n}=-me^{-\theta_{n}}
\end{equation}
where we introduced the rapidity as $k_{n}=m\sinh\theta_{n}$. The
norms of the states are related to 
\begin{equation}
\{\beta_{n}^{\dagger},\beta_{n}\}=1+\vert\gamma_{n}\vert^{2}=1+e^{-2\theta_{n}}\quad;\qquad\{\beta_{-n}^{+},\beta_{-n}\}=1+\vert\alpha_{n}\vert^{2}=1+e^{-2\theta_{n}}
\end{equation}

In order to obtain the new vacuum $\vert0,m\rangle$ we demand that
\begin{equation}
\beta_{n}\vert0,m\rangle=\beta_{-n}\vert0,m\rangle=0
\end{equation}
Such state can easily be constructed as 
\begin{equation}
\vert0,m\rangle\propto\prod_{n>0}\beta_{-n}\beta_{n}\vert0\rangle\propto\prod_{n>0}(1+\alpha_{n}b_{-n}\bar{b}_{-n})\vert0\rangle=\prod_{n>0}(1-ie^{-\theta_{n}}b_{-n}\bar{b}_{-n})\vert0\rangle
\end{equation}
We might normalize this state by dividing with $\mathcal{N}=\prod_{n>0}\sqrt{1+e^{-2\theta_{n}}}$:
\[
\vert0,m\rangle=\mathcal{N}^{-1}\prod_{n>0}(1-ie^{-\theta_{n}}b_{-n}\bar{b}_{-n})\vert0\rangle
\]

Excited states are obtained by acting with the creation operators.
They act non-trivially in their own mode numbers and that factor of
the product is modified as 
\begin{equation}
\beta_{n}^{\dagger}\vert0,m\rangle\propto b_{-n}\vert0\rangle\quad;\qquad\beta_{-n}^{\dagger}\vert0,m\rangle\propto\bar{b}_{-n}\vert0\rangle\quad;\qquad\beta_{n}^{\dagger}\beta_{-n}^{\dagger}\vert0,m\rangle\propto(e^{-\theta_{n}}+ib_{-n}\bar{b}_{-n})\vert0\rangle
\end{equation}
Thus the normalized two particle state with zero momentum and mode
number $k$ is 
\begin{equation}
\vert\{k\},m\rangle=\mathcal{N}^{-1}(e^{-\theta_{k}}+ib_{-k}\bar{b}_{-k})\prod_{n\neq k}(1-ie^{-\theta_{n}}b_{-n}\bar{b}_{-n})\vert0\rangle
\end{equation}

Let us elaborate on the groundstate energy. In the perturbed picture
the groundstate energy is 
\begin{equation}
E_{0}(L)=\sum_{n>0}\kappa_{n}=-\sum_{n>0}(\omega_{n}-k_{n})
\end{equation}
which is nothing but the sum of the zero point fluctuation energies
compared to the same expression without the perturbation. This expression
is expected to be written as 
\begin{equation}
E_{0}(L)=f_{\mathrm{bulk}}L+E_{0}^{\mathrm{TBA}}(L)=f_{\mathrm{bulk}}L-m\int\frac{d\theta}{2\pi}\cosh\theta\,\log(1+e^{-mL\cosh\theta})
\end{equation}
however $f_{\mathrm{bulk}}$ is infinite and needs regularization.
We will face a similar divergence also for the overlaps. 

\subsection{Conformal boundary conditions and overlaps}

We now turn to the description of conformal boundary conditions in
the fermionic language. The Cardy states in the Ising model can be
written in terms of the Ishibashi states as \cite{Cardy:1989fw}
\begin{eqnarray}
\vert B_{0}\rangle & = & \frac{1}{\sqrt{2}}\vert0\rangle\!\rangle+\frac{1}{\sqrt{2}}\vert{\textstyle \frac{1}{2}}\rangle\!\rangle+\frac{1}{\sqrt[4]{2}}\vert{\textstyle \frac{1}{16}}\rangle\!\rangle\nonumber \\
\vert B_{\frac{1}{2}}\rangle & = & \frac{1}{\sqrt{2}}\vert0\rangle\!\rangle+\frac{1}{\sqrt{2}}\vert{\textstyle \frac{1}{2}}\rangle\!\rangle-\frac{1}{\sqrt[4]{2}}\vert{\textstyle \frac{1}{16}}\rangle\!\rangle\nonumber \\
\vert B_{\frac{1}{16}}\rangle & = & \vert0\rangle\!\rangle-\vert{\textstyle \frac{1}{2}}\rangle\!\rangle
\end{eqnarray}
As we are focusing on the NS sector we distinguish two boundary states
$\vert B_{\pm}\rangle=$$\vert0\rangle\!\rangle\pm\vert\frac{1}{2}\rangle\!\rangle$.
Let us see how they can be described in the fermionic language. Boundary
states in the free fermion algebra are defined by 
\begin{equation}
(b_{n}\pm i\bar{b}_{-n})\vert B\rangle=0\label{eq:FBC}
\end{equation}
 and the sign reflects the fixed or the free boundary conditions.
Since the inner product matrix is diagonal in the fermionic basis
the Ishibashi states are diagonal, too. The fermionic boundary condition
also fixes how we combine the Ishibashi states of $\vert0\rangle\!\rangle$
and $\vert\frac{1}{2}\rangle\!\rangle$: 
\begin{equation}
\vert B_{\pm}\rangle=e^{\pm i\sum_{n}b_{-n}\bar{b}_{-n}}\vert0\rangle=\prod_{n}e^{\pm ib_{-n}\bar{b}_{-n}}\vert0\rangle=\prod_{n}(1\pm ib_{-n}\bar{b}_{-n})\vert0\rangle
\end{equation}
Clearly this state satisfies (\ref{eq:FBC}) and the Ishibashi condition
but it is not normalized. 

Let us note that in the UV limit, $m\to0$, the interacting vacuum
goes to the conformal vacuum $\vert0,m\rangle\to\vert0\rangle$, while
in the IR limit, $m\to\infty$ the rapidities all go to zero and $\vert0,m\rangle\to\vert B_{-}\rangle$
as was observed in \cite{Konechny:2016eek}. 

The boundary state $\vert B_{\pm}\rangle$ in the basis of the $n^{th}$
mode $\{\vert0\rangle,b_{-n}\vert0\rangle,\bar{b}_{-n}\vert0\rangle,b_{-n}\bar{b}_{-n}\vert0\rangle\}$
takes a very simple form $\{1,0,0,\pm i\}$. 

Now we are ready to test our proposal for the overlaps. Following
our suggestion the finite volume overlap of the boundary state with
the groundstate can be calculated as
\begin{equation}
g_{\pm}^{\mathrm{TCSA}}(L)=\langle0,m\vert B_{\pm}\rangle=\prod_{n}\frac{1\mp e^{-\theta_{n}}}{\sqrt{1+e^{-2\theta_{n}}}}
\end{equation}
As the boundary states are not normalizable and the boundary energies
$f_{\alpha}$ are infinite \cite{Dorey:1999cj,Dorey:2004xk} (just
as the bulk energy) a better quantity is the ratio of the groundstate
and excited state overlaps. The overlap of an excited state consisting
a pair of particles with vanishing total momentum is: 
\begin{equation}
g_{\pm}^{\mathrm{TCSA}}(\{k\},L)=\langle\{k\},m\vert B_{\pm}\rangle=\frac{e^{-\theta_{k}}\pm1}{\sqrt{1+e^{-2\theta_{k}}}}\prod_{n\neq k}\frac{1\mp e^{-\theta_{n}}}{\sqrt{1+e^{-2\theta_{n}}}}
\end{equation}
The ratio of the two quantities are 
\begin{equation}
\frac{g_{\pm}^{\mathrm{TCSA}}(\{k\}\vert L)}{g_{\pm}^{\mathrm{TCSA}}(L)}=\frac{e^{-\theta_{k}}\pm1}{1\mp e^{-\theta_{k}}}=\begin{cases}
\begin{array}{c}
\frac{e^{-\theta_{k}}+1}{1-e^{-\theta_{k}}}=\coth\frac{\theta_{k}}{2}\\
-\frac{1-e^{-\theta_{k}}}{1+e^{-\theta_{k}}}=-\tanh\frac{\theta_{k}}{2}
\end{array}\end{cases}
\end{equation}
Recall that in the block notation the reflection factor for the $+$
boundary is 
\begin{equation}
R_{+}(\theta)=\left(\frac{1}{2}\right)_{\theta}\quad;\qquad K_{+}(\theta)=R_{+}(\frac{i\pi}{2}+\theta)=i\coth\frac{\theta}{2}
\end{equation}
while for the $-$ boundary we have 
\begin{equation}
R_{-}(\theta)=\left(-\frac{1}{2}\right)_{\theta}\quad;\qquad K_{-}(\theta)=R_{-}(\frac{i\pi}{2}+\theta)=-i\tanh\frac{\theta}{2}
\end{equation}
and they agree with the exact results (\ref{eq:freeg}) up to a sign,
which can be defined into the normalization of the excited states. 

Thus we are convinced that the overlap in the massive theory can be
calculated in the TCSA framework as the matrix element of the conformal
boundary state with the massive scattering state, which is realized
on the conformal Hilbert space. 

\section{Scaling Lee-Yang model}

In the Lee-Yang model the central charge is $c=-\frac{22}{5}$ and
we have just two irreducible representations with $h=0$ and $h=-\frac{1}{5}$.
The Hilbert space is built over the two highest weight spinless fields:
$\mathbb{I}$ and the other which we denote by $\Phi$ as ${\cal H}=\mathcal{V}_{0}\otimes\bar{{\cal V}}_{0}+\mathcal{V}_{-\frac{1}{5}}\otimes\bar{{\cal V}}_{-\frac{1}{5}}$.
This is a non-unitary theory and the state with the lowest energy
is $\vert\text{vac}\rangle=\vert\Phi\rangle$. The conformal Hamiltonian
on the strip is simply 
\begin{equation}
H_{0}=\frac{2\pi}{L}(L_{0}+\bar{L}_{0}-\frac{c}{12})
\end{equation}
and we perturb it with the only relevant spinless field $\Phi$ as
(\ref{eq:HTCSA}). In order to keep the $g$-functions and the structure
constant real it is advantageous to take the normalization $\langle0\vert0\rangle=-1$
and $\langle\Phi\vert\Phi\rangle=1$. This leads to the matrix elements
\begin{equation}
\langle\Phi\vert\Phi\vert\Phi\rangle=\sqrt{\frac{2}{1+\sqrt{5}}}\frac{\Gamma(\frac{1}{5})\Gamma(\frac{6}{5})}{\Gamma(\frac{3}{5})\Gamma(\frac{4}{5})}\quad;\qquad\langle\Phi\vert\Phi\vert0\rangle=\langle0\vert\Phi\vert\Phi\rangle=1
\end{equation}
The perturbation results in a massive scattering theory of a single
particle type, whose mass is related to the coupling (\ref{eq:kappa})
as 
\begin{equation}
\kappa=2^{\frac{19}{12}}\sqrt{\pi}\frac{\left(\Gamma(\frac{3}{5})\Gamma(\frac{4}{5})\right)^{\frac{5}{12}}}{5^{\frac{5}{16}}\Gamma(\frac{2}{3})\Gamma(\frac{5}{6})}=2.642944
\end{equation}
In comparing the TCSA spectrum with those coming from TBA we note
that the bulk energy constant is 
\begin{equation}
f_{\mathrm{bulk}}=-\frac{1}{4\sqrt{3}}m^{2}
\end{equation}

\begin{figure}
\begin{centering}
\includegraphics[width=12cm]{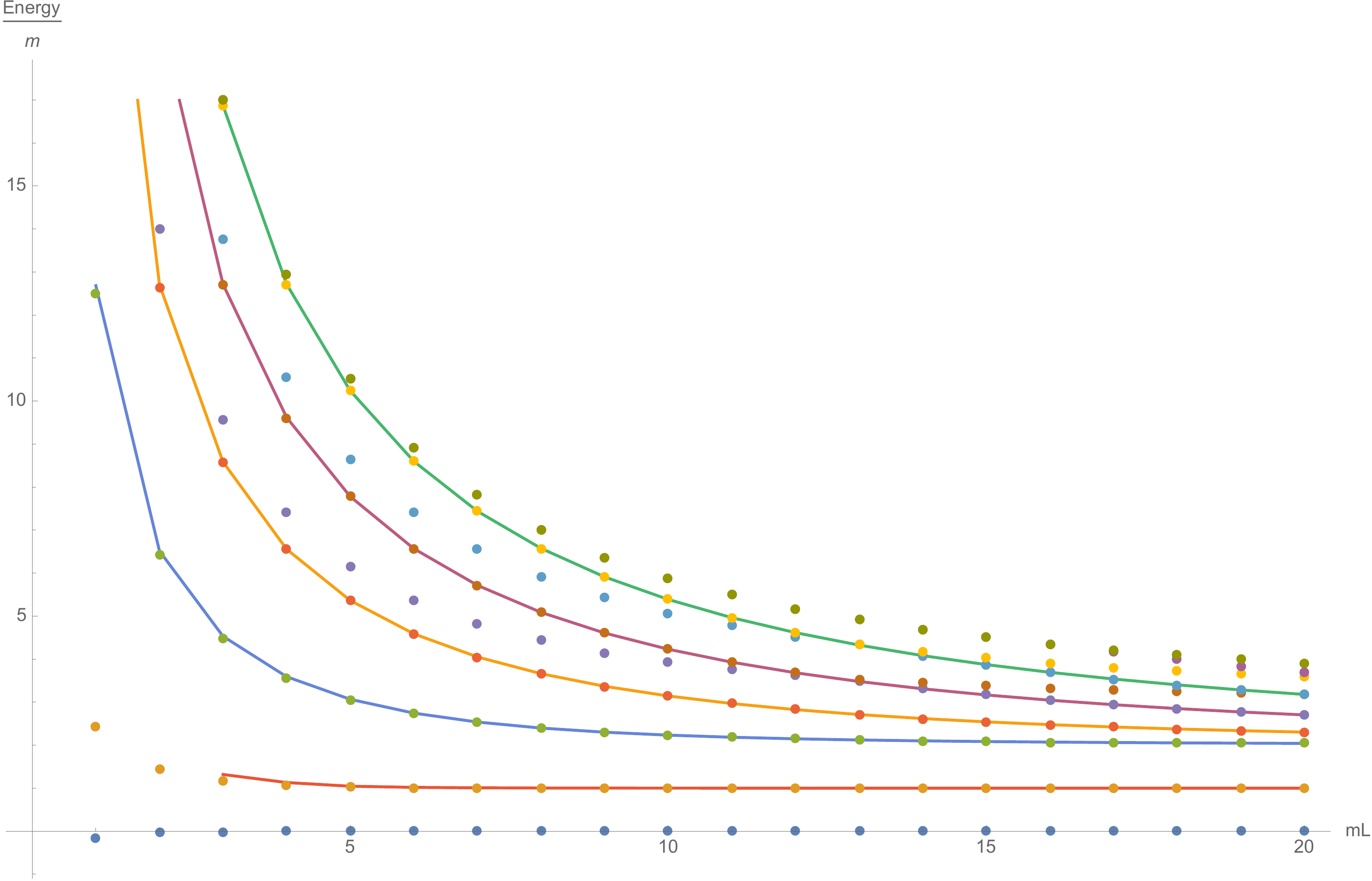}
\par\end{centering}
\caption{TCSA energy spectrum with bulk energy subtracted in the zero momentum
sector. The first state is the vacuum, the second is a standing one
particle state, while the next two are moving two-particle states.
Higher levels cross each other, for instance the fifth energy level
can be a 2- or a 3- particle scattering state depending on the volume.
Continous lines indicate 2-particle Bethe-Yang lines with quantum
numbers $n=0,\dots,4.$ }

\label{Fig:spectrum}
\end{figure}

There are two physical boundary conditions corresponding to the representations
$h=0$ and $h=-\frac{1}{5}$ labeled by $\mathbb{I}$ and $\varphi$,
respectively. The related Cardy states can be expressed in terms of
the Ishibashi states as \cite{Dorey:1997yg}
\begin{align}
\vert B_{\mathbb{I}}\rangle & =-\left(\frac{1}{2}+\frac{1}{2\sqrt{5}}\right)^{\frac{1}{4}}\vert0\rangle\!\rangle+\left(\frac{1}{2}-\frac{1}{2\sqrt{5}}\right)^{\frac{1}{4}}\vert-{\textstyle \frac{1}{5}}\rangle\!\rangle\label{eq:LYBstate}\\
\vert B_{\varphi}\rangle & =\left(1-\frac{2}{\sqrt{5}}\right)^{\frac{1}{4}}\vert0\rangle\!\rangle+\left(1+\frac{2}{\sqrt{5}}\right)^{\frac{1}{4}}\vert-{\textstyle \frac{1}{5}}\rangle\!\rangle
\end{align}
The corresponding reflection factors are (\ref{eq:Rid},\ref{eq:Rphi}),
see \cite{Dorey:1997yg}, while the boundary energies are 
\begin{equation}
f_{\mathbb{I}}=\frac{1}{4}(\sqrt{3}-1)m\quad;\qquad f_{\varphi}=f_{\mathbb{I}}-m\sin\frac{\pi}{12}\label{eq:bdryenLY}
\end{equation}

In implementing TCSA for the Lee-Yang model we generated the periodic
Hilbert space (\ref{eq:Hspace}) with energy cuts ranging from $10$
to $18$ in the chiral Verma modules and diagonalized the truncated
Hamiltonian (\ref{eq:HTCSA}) numerically. The low lying spectrum
with the bulk energy subtracted is presented on Figure (\ref{Fig:spectrum}).
The two particle Bethe-Yang lines 
\begin{equation}
Q(\theta_{n})=mL\sinh\theta_{n}-i\log S(2\theta_{n})=2\pi n\quad;\qquad E_{n}(L)=2m\cosh\theta_{n}\label{eq:BY2}
\end{equation}
are plotted for $n=0,\dots,4$. The state with $n=0$ is actually
a 1-particle state, which can be interpreted as a boundstate for $mL>3$.
Having obtained the spectrum together with the corresponding eigenvalues
we calculated their overlaps with the boundary states (\ref{eq:LYBstate})
constructed on the truncated Hilbert space. In the following we report
on the comparison of the outcome of these TCSA overlap calculations
(\ref{eq:tcsag0}), (\ref{eq:tcsagex}) against various available
results for small (UV), large (IR) and intermediate (TBA) volumes.
We investigate in details the first four eigenstates, which includes
the finite volume vacuum, a 1-particle state and two 2-particle states. 

\subsection{Small volume checks of TCSA overlaps}

For small volumes we can compare the TCSA results with CPT calculations.
The groundstate $g$-function has the expansion 
\begin{equation}
\log g_{\mathbb{I}}^{\mathrm{CFT}}=\log\,_{L}\langle\text{vac},\lambda\vert B_{\mathbb{I}}\rangle=\frac{1}{4}\log\left(\frac{1}{2}+\frac{1}{2\sqrt{5}}\right)+\sum_{i=1}^{\infty}d_{i}(mL^{\frac{12}{5}}/\kappa)^{i}\label{eq:TCSAvspert}
\end{equation}
where conformal perturbation theory gives the following coefficients
\cite{Dorey:1999cj,Dorey:2004xk}:
\begin{equation}
d_{1}=-0.25312\quad;\qquad d_{2}=0.0775\quad;\qquad d_{3}=-0.0360\quad;\qquad d_{4}=0.0195
\end{equation}

\begin{figure}[H]
\begin{centering}
\includegraphics[width=7cm]{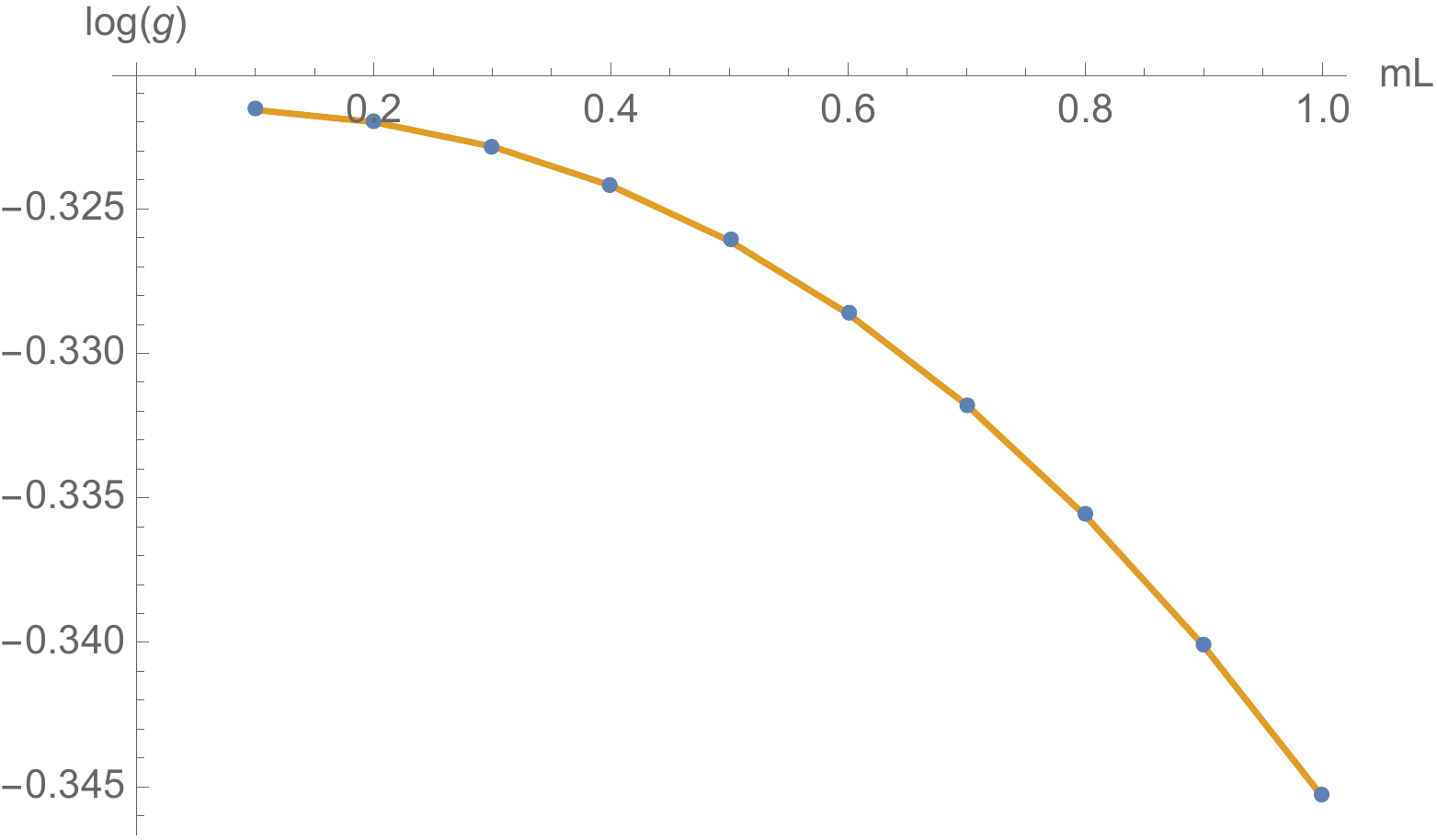}~~~~\includegraphics[width=7cm]{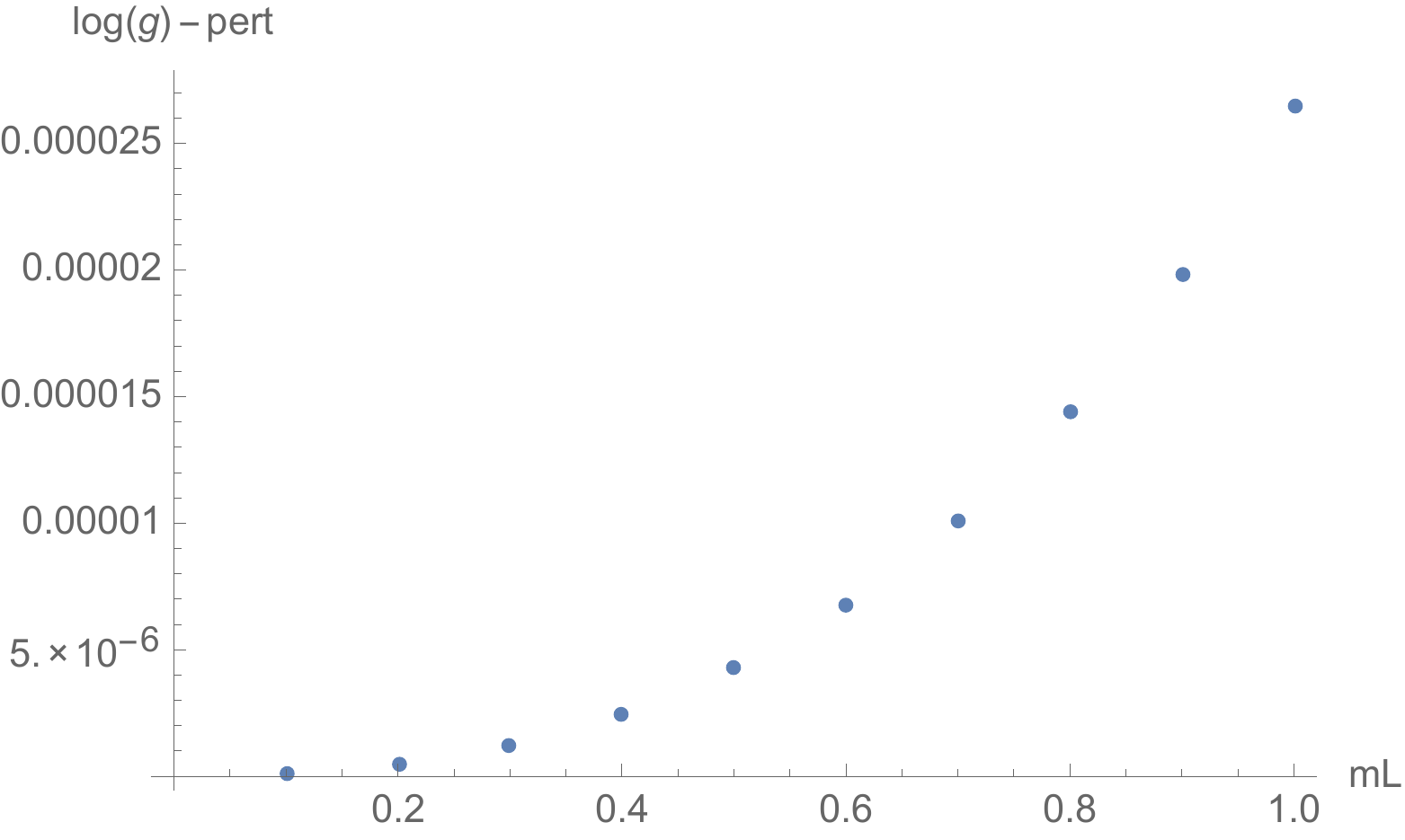}
\par\end{centering}
\caption{Comparison of TCSA and conformal perturbation theory (\ref{eq:TCSAvspert}).
On the left: TCSA data points for $\log(g_{\mathbb{I}}^{\mathrm{CFT}})$
are plotted against the dimensionless volume $mL=0.1,\dots,1$ with
blue dots and conformal perturbation theory up to $d_{4}$ with a
continous line. The difference of the two is shown on the right, indicating
the next order polynomial correction. }

\label{Fig:UV}
\end{figure}

In Figure (\ref{Fig:UV}) we compare the TCSA results with the CPT
calculations and find complete agreement. This confirms that the coefficients
and signs coming from various square-roots in the boundary state are
implemented correctly\footnote{Note that some of our signs in the boundary state are different from
\cite{Dorey:1999cj}, which can be related to different conventions. } and gives us confidence to move on to the large volume analysis. 

\subsection{Large volume checks of TCSA overlaps}

\begin{figure}[H]
\begin{centering}
\includegraphics[width=7cm]{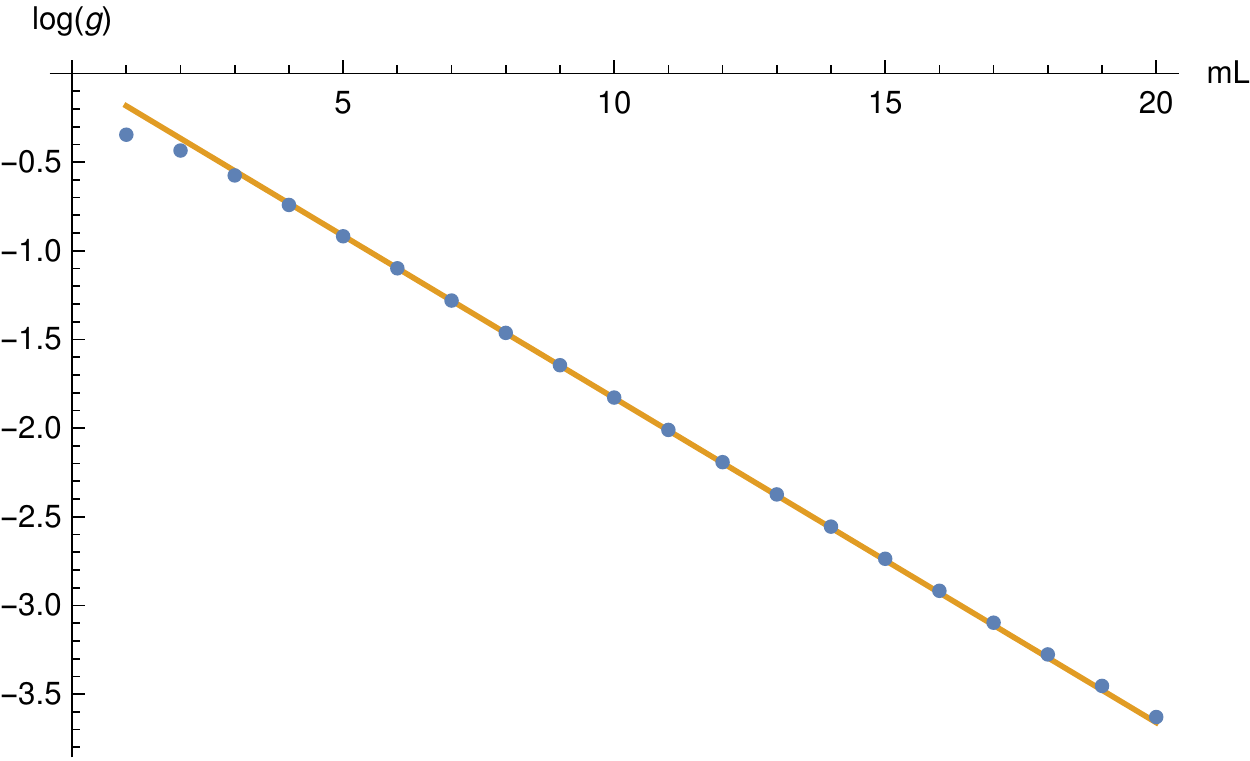}~~~~\includegraphics[width=7cm]{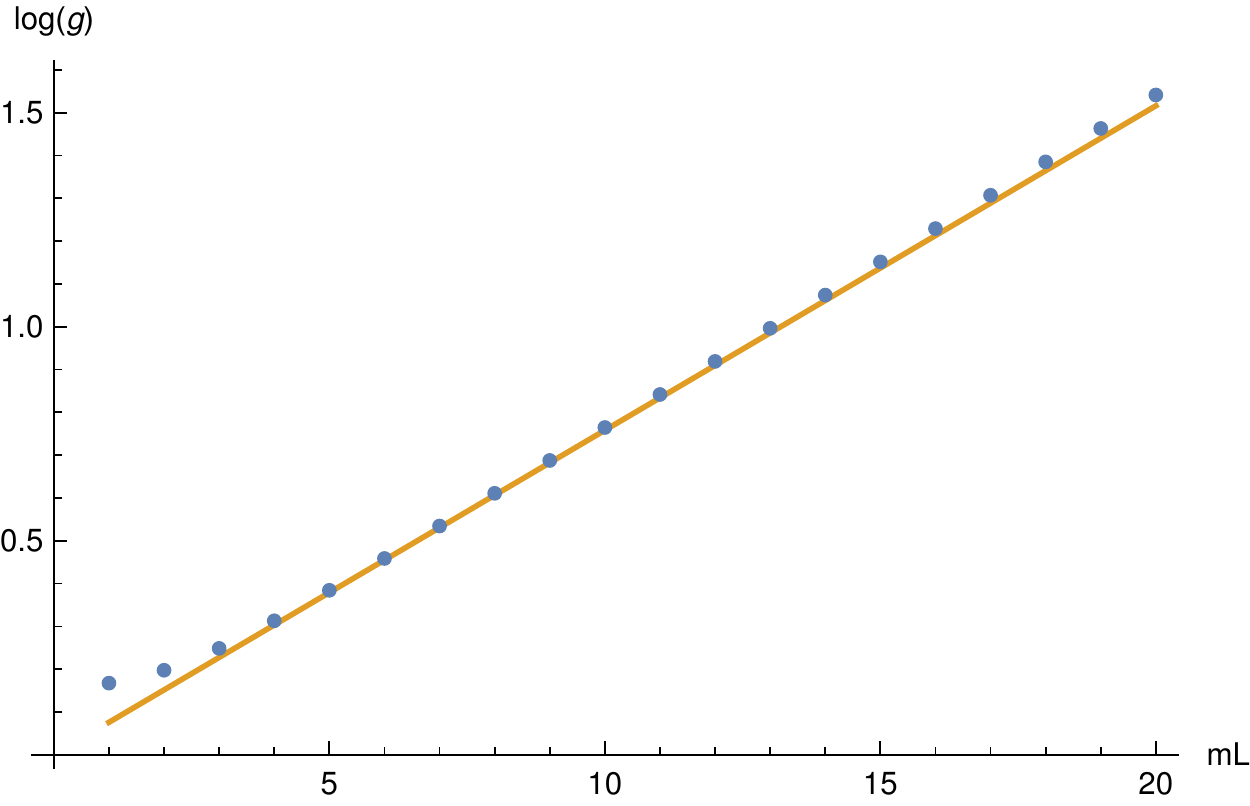}
\par\end{centering}
\caption{The overlap of the vacuum with the boundary states. On the left, logarithm
of the overlap between the vacuum and the identity boundary state
(blue dots) is plotted against the dimensionless volume together with
$-f_{\mathbb{I}}mL$ (continuous line). On the right the overlap with
the $\varphi$ boundary and $-f_{\varphi}mL$. }

\label{Fig:gidvraw}
\end{figure}

\begin{figure}[H]
\begin{centering}
\includegraphics[width=7.5cm]{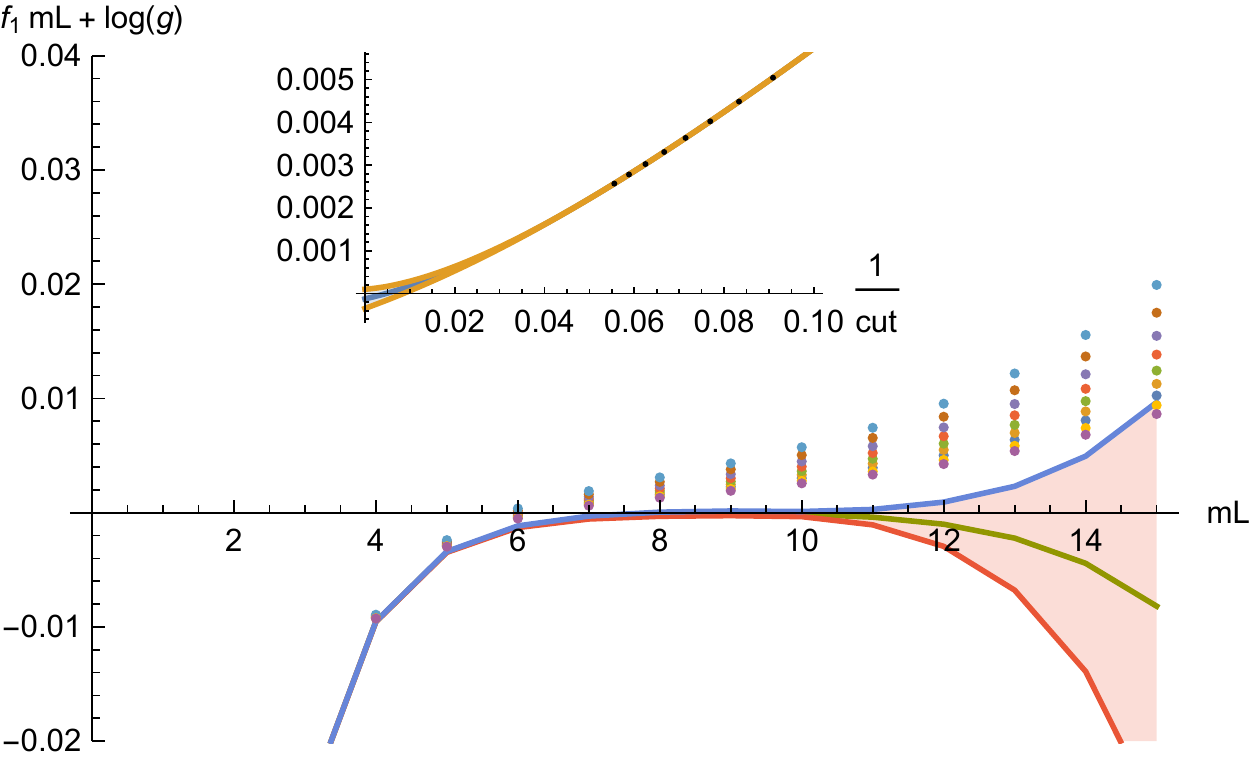}~~~~\includegraphics[width=7.5cm]{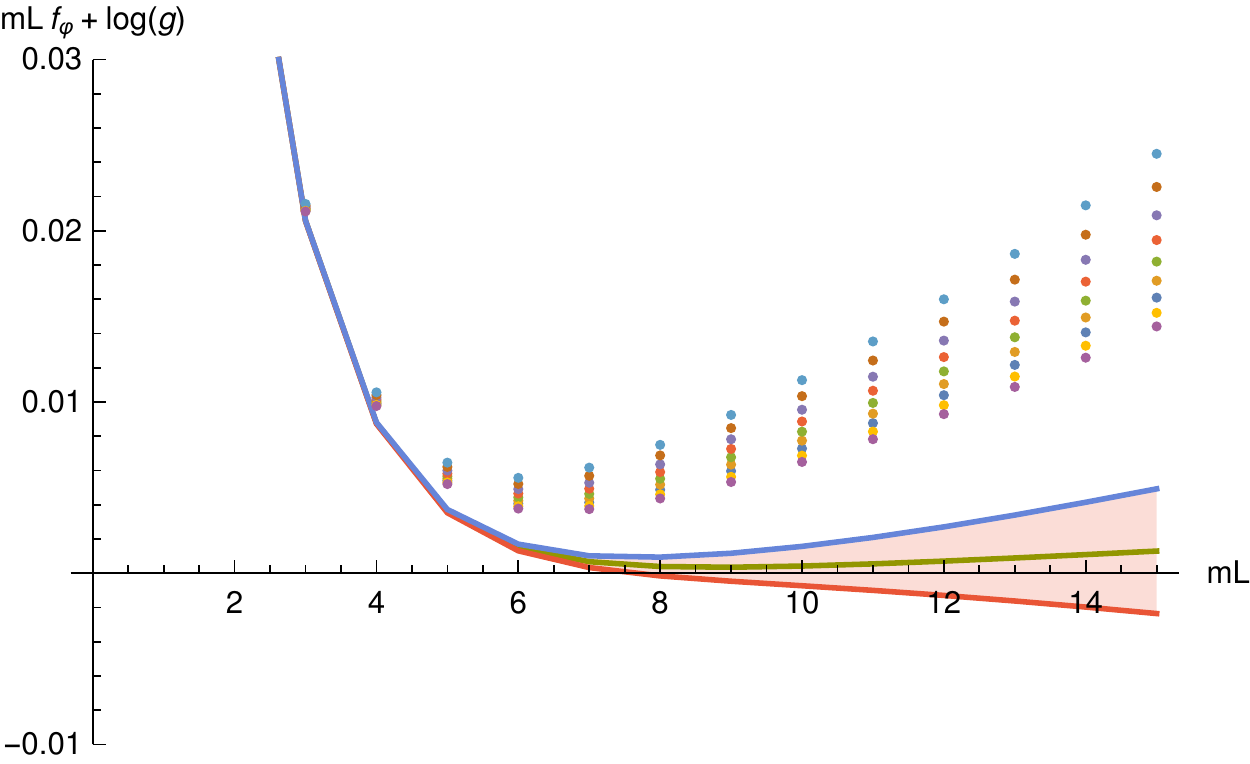}
\par\end{centering}
\caption{The difference between the logarithm of the overlap and the boundary
energy, $\log g_{\mathbb{I}}^{\mathrm{CFT}}+f_{\mathbb{I}}mL$, for
various cuts ranging from $10$ to $18$ on the left. TCSA data are
plotted with dots, such that higher cuts are closer to zero. Extrapolated
results with $90\%$ confidence range are plotted with solid lines.
The inset shows the extrapolation in the cut for $mL=10$. Similar
plots for the $\varphi$ boundary is on the right. }

\label{Fig:extrapolation}
\end{figure}

Let us start with the vacuum overlaps. On Figure (\ref{Fig:gidvraw})
the logarithm of (\ref{eq:tcsag0}) is plotted as a function of the
dimensionless volume $mL$ for both boundary conditions. Clearly the
TCSA overlaps (\ref{eq:logf}) contain the boundary energies (\ref{eq:bdryenLY}).
In the following we subtract these linear contributions from the logarithm
of the overlaps and analyze only the difference. In doing so we can
observe stronger dependence on the cut. We then introduce an (conservative)
extrapolation for each volume in the cut of the form
\begin{equation}
(g_{\mathbb{I}}^{\mathrm{CFT}})\vert_{\mathrm{cut}}=(g_{\mathbb{I}}^{\mathrm{CFT}})\vert_{\mathrm{\infty}}+\frac{b}{\mathrm{cut}}+\frac{c}{\mathrm{cut}^{2}}+\frac{d}{\mathrm{cut}^{\frac{12}{5}}}+\frac{e}{\mathrm{cut}^{3}}\label{eq:extrapolation}
\end{equation}
where \emph{cut }is the truncation level of the chiral Virasoro Verma
modules. The extrapolated results, presented on Fig \ref{Fig:extrapolation},
show both that the extrapolation is correct and that the logarithm
of the TBA normalized g-functions vanish for large volumes. In the
next subsection we compare these extrapolated results to TBA calculations. 

We next analyze the overlaps with a standing one particle state. Since
both reflection factors have a pole at $i\pi/2$ with residues
\begin{equation}
R_{\alpha}(\theta)=i\frac{g_{\alpha}^{2}}{2\theta-i\pi}+\dots\quad;\qquad g_{\mathbb{I}}=2\sqrt{2\sqrt{3}-3}\quad;\quad g_{\varphi}=\sqrt{2}\sqrt[4]{3}\sin^{2}\left(\frac{\pi}{8}\right)\csc^{2}\left(\frac{5\pi}{24}\right)
\end{equation}
the leading finite size correction takes the form (\ref{eq:Kasodd}):
\begin{equation}
\log\,_{L}\langle1,\lambda\vert B_{\alpha}\rangle+f_{\alpha}mL=\frac{g_{\alpha}}{2}\sqrt{\bar{\rho}_{1}^{+}}+O(e^{-mL})=\frac{g_{\alpha}}{2}\sqrt{mL}+O(e^{-mL})
\end{equation}

\begin{figure}[H]
\begin{centering}
\includegraphics[width=12cm]{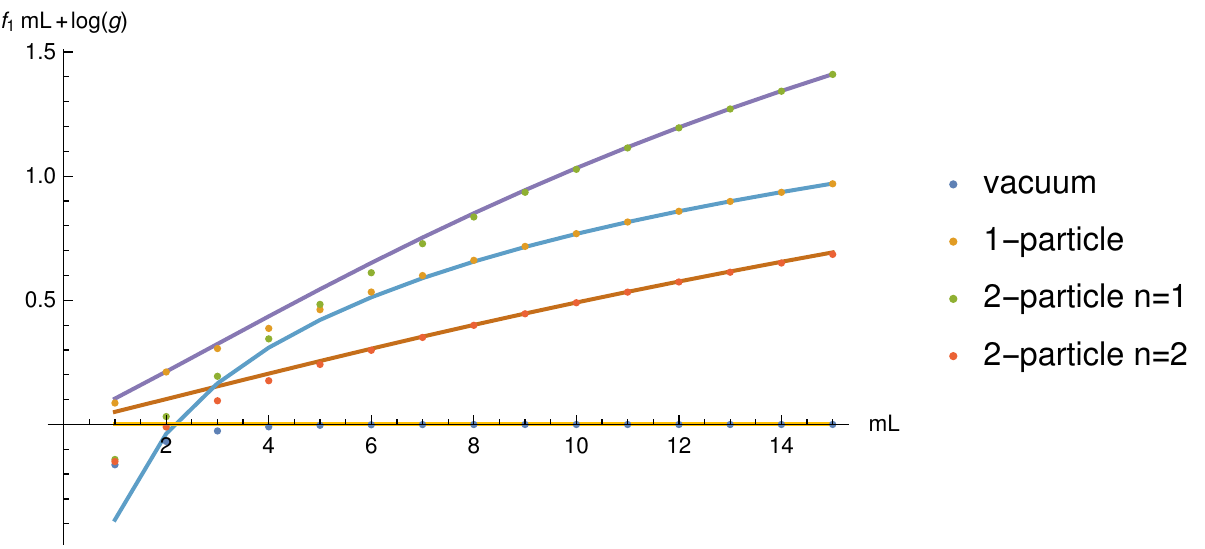}
\par\end{centering}
\caption{The expression $\log\,_{L}\langle n,\lambda\vert B_{\mathbb{I}}\rangle+f_{\mathbb{I}}mL$
is plotted for the first four states with dots. These includes the
vacuum, the 1-particle state and two 2-particle states with quantization
numbers $n=1,2$ in (\ref{eq:BY2}). The asymptotic expressions are
plotted with solid lines. The extrapolation errors are negligible
on the plot. }

\label{loggid}
\end{figure}

\begin{figure}[H]
\begin{centering}
\includegraphics[width=12cm]{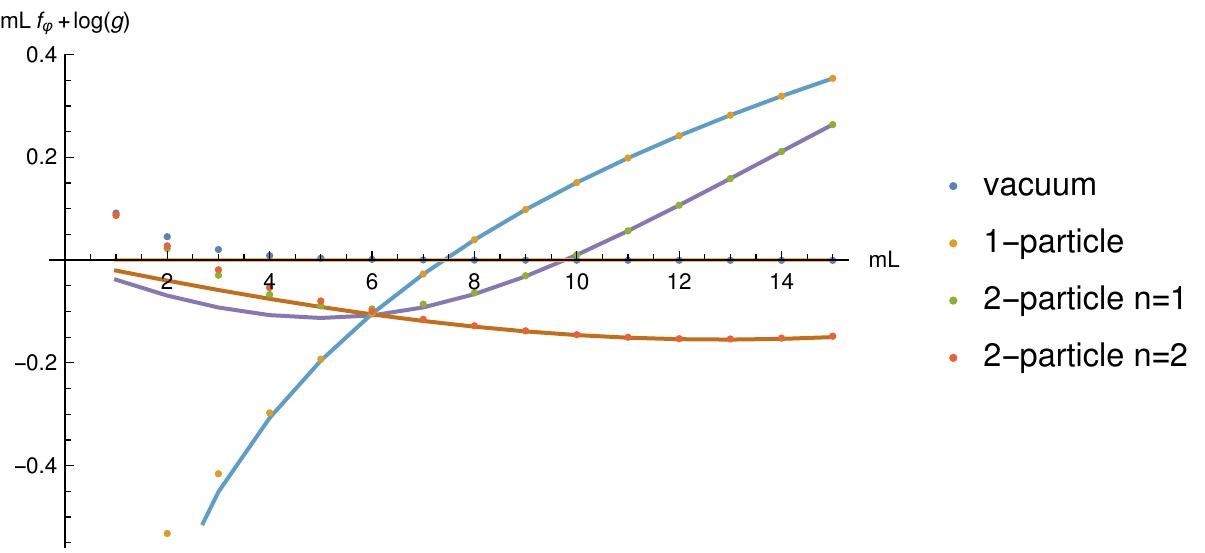}
\par\end{centering}
\caption{The expression $\log\,_{L}\langle n,\lambda\vert B_{\mathbb{\varphi}}\rangle+f_{\mathbb{\varphi}}mL$
is compared for the first four states including the vacuum, the 1-particle
state and two 2-particle states with quantization numbers $n=1,2$
to the asymptotic expressions (solid lines). }

\label{loggph}
\end{figure}

The asymptotic form of the overlaps with two particle states takes
the form
\begin{equation}
\log\,_{L}\langle2,\lambda\vert B_{\alpha}\rangle+f_{\alpha}mL=\frac{K_{\alpha}(\theta)}{\sqrt{S(2\theta)}}\sqrt{\frac{\rho_{1}^{+}(\theta)}{\rho_{1}^{-}(\theta)}}+O(e^{-mL})=\frac{R_{\alpha}(\frac{i\pi}{2}-\theta)}{\sqrt{S(2\theta)}}\sqrt{\frac{p'(\theta)L}{p'(\theta)L+2\phi(2\theta)}}+O(e^{-mL})
\end{equation}
where for any volume $L$ the rapidity $\theta$ is determined from
the quantization condition (\ref{eq:BY2}). The TCSA overlap calculations
for the first 4 states are compared to the asymptotic expressions
in Figures (\ref{loggid}) and (\ref{loggph}). Clearly we find a
convincing evidence. 


\subsection{TBA checks against TCSA at intermediate volume }

\begin{figure}[H]
\begin{centering}
\includegraphics[width=7.5cm]{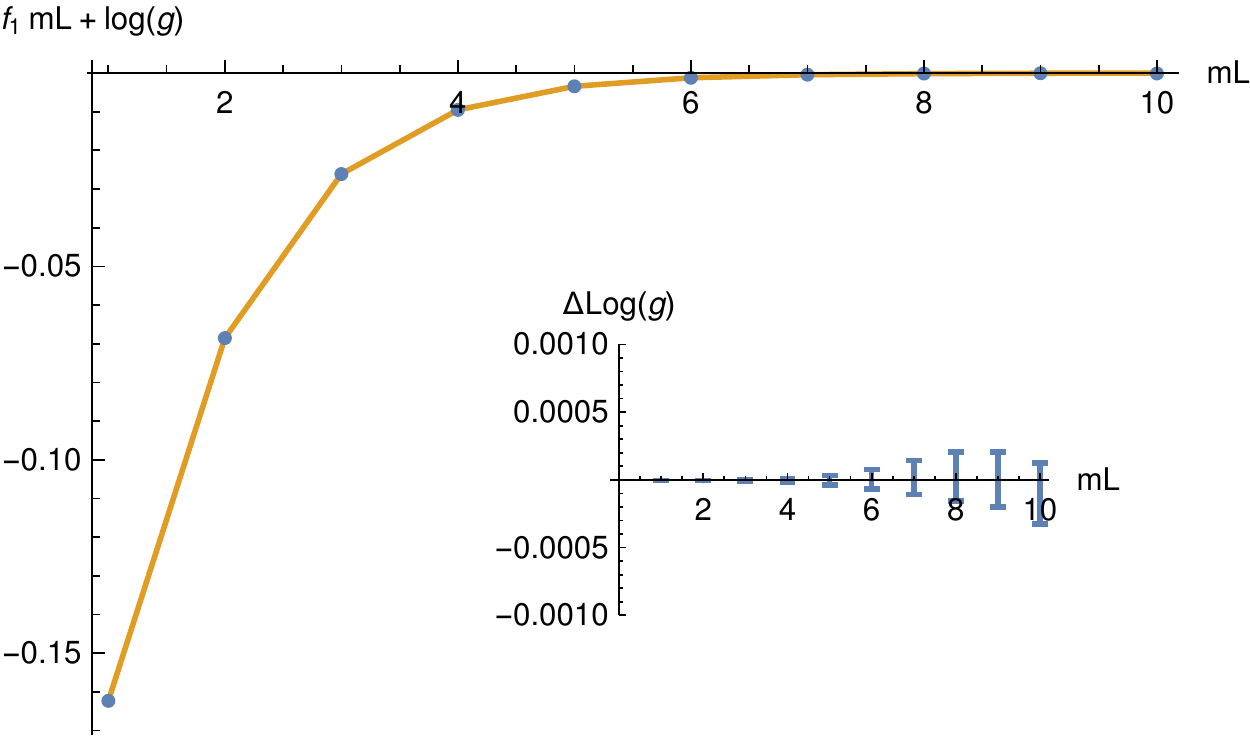}~~~~\includegraphics[width=7.5cm]{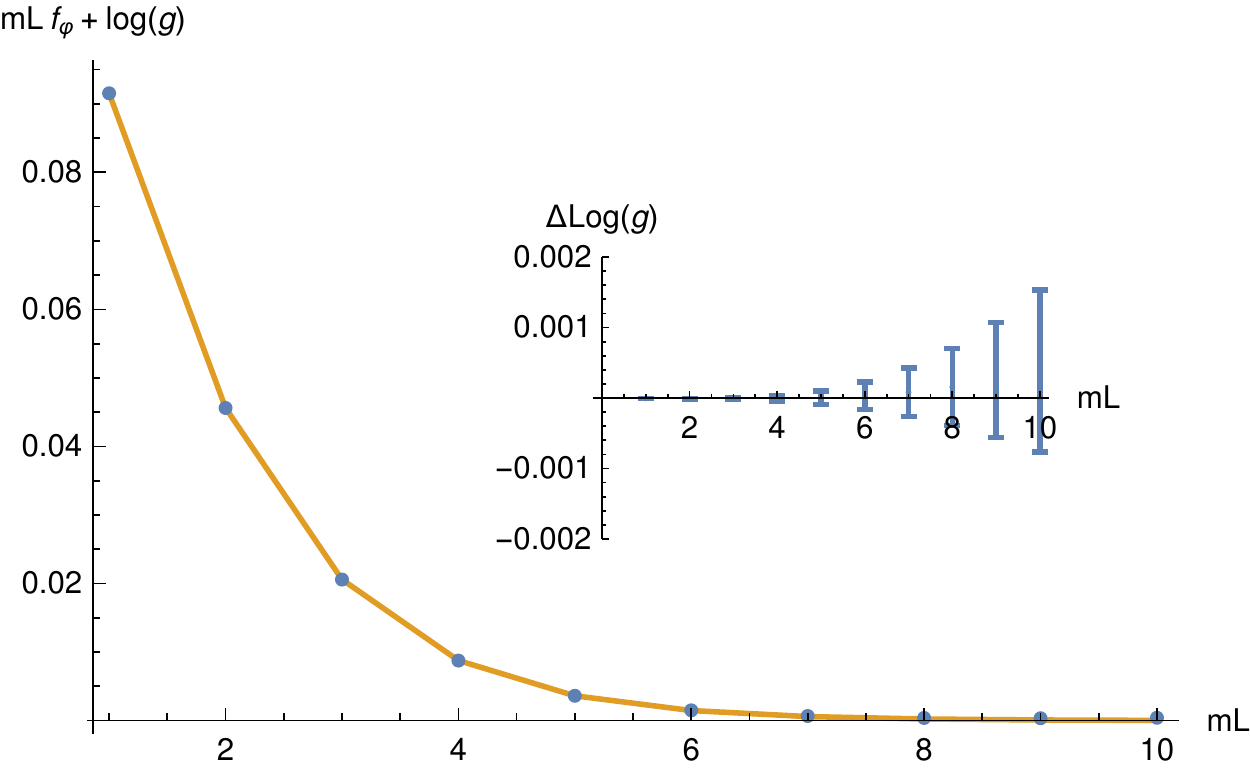}
\par\end{centering}
\caption{The logarithm of the groundstate g-functions calculated numerically
from TBA equations are plotted with solid lines against extrapolated
TCSA with dots for the identity boundary condition on the left while
for $\varphi$ on the right. The insets enlarge the differences of
the two. }

\label{Fig:tbav}
\end{figure}

Finally, we compare the TCSA data against the solutions of the TBA
equations. We start with the groundstate expression. 
\begin{equation}
g_{\mathbb{I}}(L)=\sqrt{\frac{\det(1-K^{-})}{\det(1-K^{+})}}\exp\left\{ \int\frac{d\theta}{4\pi}\left(\phi_{\mathbb{I}}(\theta)-\phi(2\theta)-\pi\delta(\theta)\right)\log(1+e^{-\epsilon(\theta)})\right\} 
\end{equation}
We explain in Appendix B how to evaluate this expression numerically.
The comparison is shown on the left of Figure \ref{Fig:tbav}. 

Interestingly the $\varphi$ boundary has a boundary boundstate so
the g-function takes the form \cite{Dorey:1999cj}
\begin{equation}
g_{\varphi}(L)=\sqrt{\frac{\det(1-K^{-})}{\det(1-K^{+})}}(1+e^{-\epsilon(i\pi/12)})\exp\left\{ \int\frac{d\theta}{4\pi}\left(\phi_{\varphi}(\theta)-\phi(2\theta)-\pi\delta(\theta)\right)\log(1+e^{-\epsilon(\theta)})\right\} 
\end{equation}
This is confirmed by comparing the TBA results to TCSA on the right
of Figure \ref{Fig:tbav}. 

A standing one particle state is obtained by analytical continuation
and takes the following form for the identity boundary 
\begin{equation}
g_{\mathbb{I}}(\{0\},L)=\sqrt{\frac{\det(1-K_{\mathrm{ex}}^{-})}{\det(1-K_{\mathrm{ex}}^{+})}}\frac{R_{\mathbb{I}}(\bar{\theta}_{1})}{\sqrt{S(2\bar{\theta}_{1})}}\exp\left\{ \int\frac{d\theta}{4\pi}\left(\phi_{\mathbb{I}}(\theta)-\phi(2\theta)-\pi\delta(\theta)\right)\log(1+e^{-\epsilon(\theta)})\right\} 
\end{equation}
In Appendix B we spell out the details how we calculated all the ingredients
of this expression numerically. In the case of the $\varphi$ boundary
we have to include additionally the contribution of the boundary boundstate:
\begin{align}
g_{\varphi}(\{0\},L) & =\sqrt{\frac{\det(1-K_{\mathrm{ex}}^{-})}{\det(1-K_{\mathrm{ex}}^{+})}}\frac{R_{\varphi}(\bar{\theta}_{1})}{\sqrt{S(2\bar{\theta}_{1})}}(1+e^{-\epsilon(i\pi/12)})\times\nonumber \\
 & \qquad\qquad\exp\left\{ \int\frac{d\theta}{4\pi}\left(\phi_{\varphi}(\theta)-\phi(2\theta)-\pi\delta(\theta)\right)\log(1+e^{-\epsilon(\theta)})\right\} 
\end{align}

\begin{figure}[H]
\begin{centering}
\includegraphics[width=7.5cm]{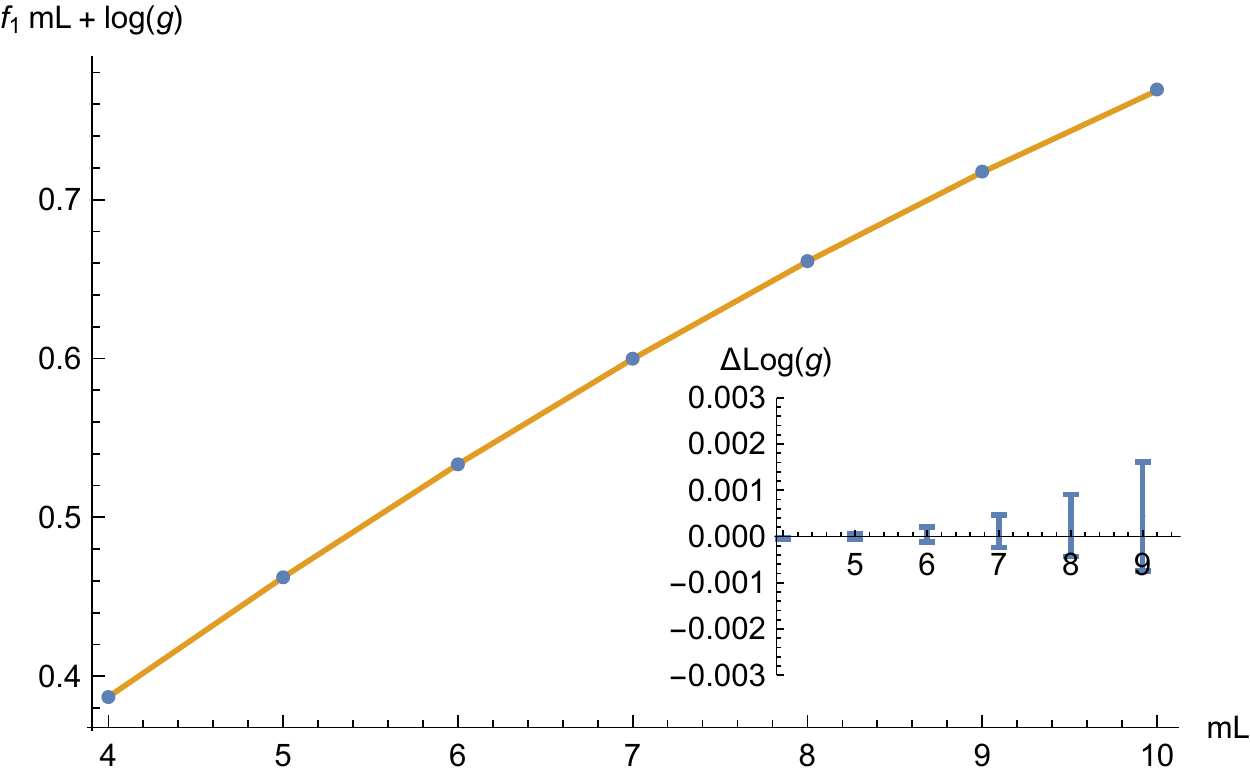}~~~~\includegraphics[width=7.5cm]{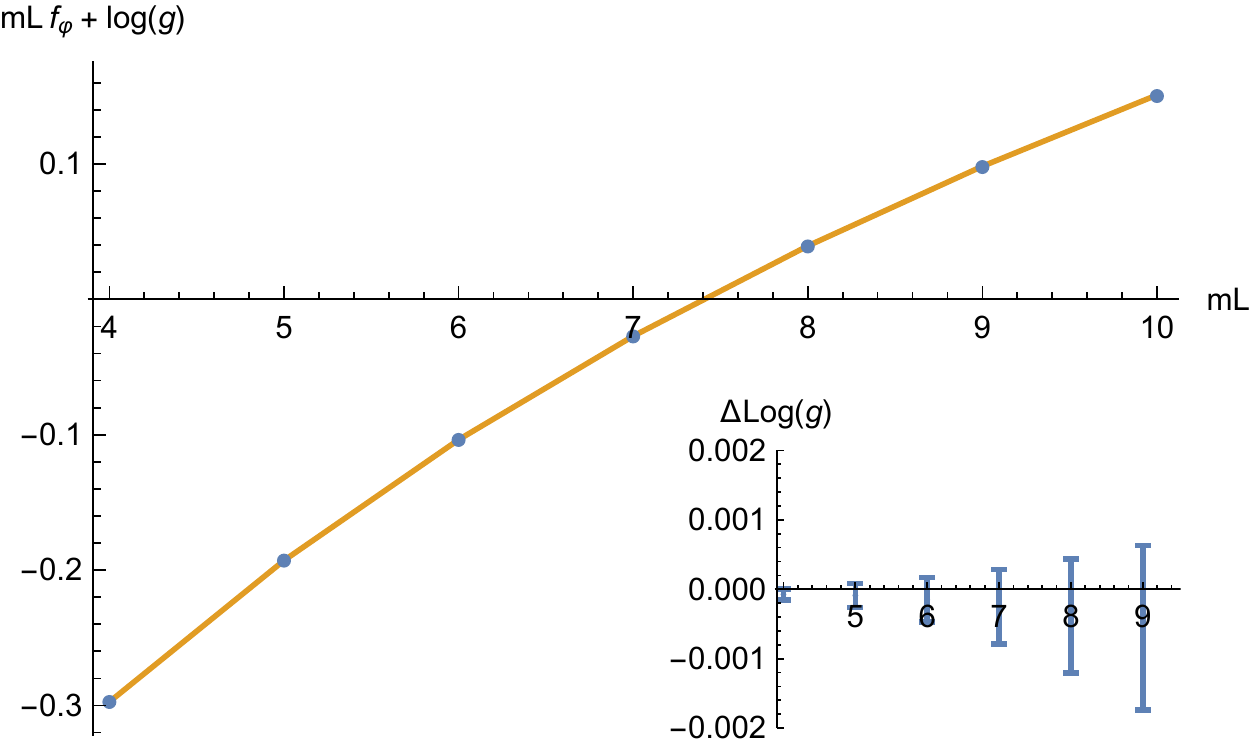}
\par\end{centering}
\caption{The 1-particle g-functions calculated numerically from TBA equations
are plotted with solid lines against extrapolated TCSA with dots for
the identity boundary condition on the left while for $\varphi$ on
the right. }

\label{Fig:tba1p}
\end{figure}

The comparison of these two cases are shown on Figure \ref{Fig:tba1p}.
We note that the excited state TBA equation has a solution only for
$mL>3$ in the form of (\ref{eq:tbaex}). 

In summarizing, we can say that in all the investigated cases we found
convincing agreement between the TBA, TCSA and CPT results.

\section{Conclusions}

The aim of our paper was to elaborate on the finite volume boundary
state. In doing so we first provided a simple argument to explain
the structure of the asymptotic overlaps. In order to focus on the
conceptual issues we presented the ideas for the simplest integrable
theories having a single particle type only. We showed that neglecting
exponentially small finite size corrections the overlap contains the
product of infinite volume overlaps/reflection factors and the square
root of the ratio of two determinants originating from the normalization
of finite volume states. We then analyzed carefully the analytical
continuation of the ground-state g-function in the scaling Lee-Yang
theory and obtained excited state g-functions. In order to test these
formulas we developed a new way of calculating finite volume overlaps
in TCSA. Assuming the absence of boundary perturbations in the Lagrangian
description it is simply the overlap of the conformal boundary state
with the eigenstate of the TCSA Hamiltonian. We tested this idea with
explicit calculations in the thermally perturbed Ising model aka the
free massive fermion. In the scaling Lee-Yang model our results confirm
completely the excited state g-functions obtained by analytical continuation.
In particular for the $\varphi$ boundary, which has a boundary boundstate,
we confirm \cite{Dorey:1999cj} that the g-function has a factor signaling
this boundstate. Additionally we presented a numerical implementation
for the solutions of the various TBA equations and the calculations
of the overlaps. 

Thorough the paper we assumed that the boundary is not perturbed only
the bulk. It would be very tempting to extend the analysis for boundary
perturbations. In doing so the bulk-boundary operator product expansion
could be used to represent the boundary perturbations in the bulk
as an operator dressing the conformal boundary state. 

We chose the Lee-Yang model to test our ideas as the implementation
of the TCSA method is probably the simplest. Recent developments however
in TCSA make it possible to apply the method for the sinh-Gordon theory
\cite{Bajnok:2019cdf,Konik:2020gdi} where the extension of the TBA
results \cite{Caetano:2020dyp} for excited states could be checked. 

Here we analyzed only diagonally scattering theories. It would very
interesting to extend and test these techniques for non-diagonal scatterings.
Then results such as \cite{Kostov:2019fvw} could be tested. Also
challenging is the calculation of the g-functions for the $O(N)$
models, where results from \cite{Gombor:2017qsy} can be relevant. 

\subsection*{Acknowledgments}

We thank Gerard Watts, Anatoly Konechny, Stephan Fredenhagen, Maté
Lencsés, Balázs Pozsgay, Márton Kormos, Gábor Takács and Tamás Gombor
for the useful discussions and comments, the NKFIH research Grants
K116505, K134946 and UK\_Gyak for supports.

\appendix

\section{The generic finite volume boundary state \label{sec:GenBstate}}

The finite volume boundary state can be expressed in the basis of
the finite volume Hilbert space with periodic boundary condition in
the form: 
\begin{equation}
\vert B\rangle_{L}=\vert B\rangle_{L}^{\mathrm{even}}+\vert B\rangle_{L}^{\mathrm{odd}}
\end{equation}
where we already calculated the even part in the bulk of the paper.
In the following we assume that $\mathrm{res}_{\theta=0}K(\theta)\neq0$
and calculate the odd part, which contains a standing particle
\begin{equation}
\vert B\rangle_{L}^{\mathrm{odd}}=\sum_{N}\sum_{\{n_{j}\}}K_{N}^{\mathrm{odd}}(0,n_{1},\dots,n_{N})_{L}\vert n_{1},\dots,n_{N},0,-n_{N},\dots,-n_{1}\rangle_{L}
\end{equation}
Similarly how we related the resolution of the identity for large
volumes we can also replace the summation for pairs of particles for
integration of those pairs. In doing so we have to use the relation
\begin{equation}
Q_{j}^{\mathrm{odd}}=p(\theta_{j})L-i\log S(\theta_{j})-i\sum_{k:k\neq j}\log S(\theta_{j}-\theta_{k})-i\sum_{k}\log S(\theta_{j}+\theta_{k})=2\pi n_{j}
\end{equation}
and change variable from $\{n_{j}\}$ to $\{\theta_{j}\}$. The corresponding
Jacobian is $\rho_{N}^{\mathrm{odd}}=\mbox{\ensuremath{\det\left|\frac{\partial Q_{j}^{\mathrm{odd}}}{\partial\theta_{i}}\right|}}$.
Additionally, we also express the finite volume state in terms of
the infinite volume state and arrive at 
\begin{equation}
\vert B\rangle_{L}^{\mathrm{odd}}=\sum_{N}\prod_{i}\int\frac{d\theta_{i}}{4\pi}K_{N}^{\mathrm{odd}}(0,\theta_{1},\dots,\theta_{N})_{L}\frac{\rho_{N}^{\mathrm{odd}}}{\sqrt{\prod_{i}S(-2\theta_{i})\rho_{2N+1}}}\vert0,-\theta_{1},\theta_{1},-\theta_{2},\theta_{2},\dots,-\theta_{N},\theta_{N}\rangle
\end{equation}
In calculating $\rho_{2N+1}$ for states of the form $\{\theta_{1},\dots,\theta_{N},0,-\theta_{1},\dots,-\theta_{N}\}$
we have the following factorization \cite{Bajnok:2020xoz}
\begin{equation}
\rho_{2N+1}=\mbox{det}\left|\begin{array}{ccc}
A & b & B\\
b & a & b\\
B & b & A
\end{array}\right|=\mbox{det}\left|\begin{array}{ccc}
A-B & 0 & 0\\
b & a & 2b\\
B & b & A+B
\end{array}\right|=\rho_{N}^{\mathrm{odd}}\bar{\rho}_{N+1}^{+}
\end{equation}
where $a=mL+2\sum_{j=1}^{N}\phi_{j}$ with $\phi_{j}=\phi(\theta_{j})$
while the $j^{th}$ component of the vector $b$ is $b_{j}=-\phi_{j}$.
For completeness we spell out that 
\begin{equation}
A_{ij}=\delta_{ij}(p_{i}'L+\phi_{j}+\sum_{k}\phi_{ik})-\phi_{ij}\quad;\qquad B_{ij}=-\phi(\theta_{i}+\theta_{j})
\end{equation}
with $\phi_{ij}=\phi(\theta_{i}-\theta_{j})$, thus one can easily
see that $\text{det\ensuremath{\vert A-B\vert=\rho_{N}^{\mathrm{odd}}}. }$
By comparing this expression with the infinite volume boundary state
we arrive at the relation
\begin{equation}
K_{N}^{\mathrm{odd}}(\theta_{1},\dots,\theta_{N})_{L}=\frac{\sqrt{\rho_{2N+1}}}{\sqrt{\prod_{i}S(2\theta_{i})}\rho_{N}^{\mathrm{odd}}}\frac{g}{2}\prod_{i}K_{2}(\theta_{i})+O(e^{-mL})=\frac{g}{2}\prod_{i}\frac{K_{2}(\theta_{i})}{\sqrt{S(2\theta_{i})}}\sqrt{\frac{\bar{\rho}_{N+1}^{+}}{\rho_{N}^{\mathrm{odd}}}}+O(e^{-mL})\label{eq:Kasodd}
\end{equation}
The first few terms of this expression agrees with \cite{Horvath:2018gat}. 

\section{Numerical solution of the TBA equations and overlaps\label{sec:numTBA}}

In this Appendix we explain how we solve numerically the TBA equations
and calculate the g-functions. 

\subsection{Groundstate}

In the following we present a numerical approach to the overlaps.
The real $\theta$ line will be discretized into $N$ points $\theta_{i}$
with distance $\theta_{i+1}-\theta_{i}=d\theta$, such that the label
of the origin is $i0$: $\theta_{i0}=0$. Each function $f(\theta)$
is then represented by a vector $f_{i}=f(\theta_{i})$ while the kernel
in the integration by a matrix 
\begin{equation}
\phi_{ij}=\frac{d\theta}{2\pi}\phi(\theta_{i}-\theta_{j})
\end{equation}
The groundstate TBA equation is a vector equation of the form 
\begin{equation}
\epsilon_{i}=mL\cosh\theta_{i}-\sum_{j}\phi_{ij}\log(1+e^{-\epsilon_{j}})
\end{equation}
This can be solved by Findroot in Mathematica or by iterations. The
groundstate energy is 
\begin{equation}
E_{0}(L)/m=-\frac{d\theta}{2\pi}\sum_{i}\cosh\theta_{i}\log(1+e^{-\epsilon_{i}})
\end{equation}
For the calculation of the g-function we introduce the matrices 
\begin{equation}
K_{ij}^{\pm}=\frac{d\theta}{2\pi}\frac{\phi(\theta_{i}-\theta_{j})\pm\phi(\theta_{i}+\theta_{j})}{1+e^{\epsilon_{i}}}
\end{equation}
 for indices when both $\theta_{i},\theta_{j}>0$. The saddle point
value of the g-function is 
\begin{align}
\log g_{\alpha}= & -\frac{1}{4}\log(1+e^{-\epsilon_{i0}})+\frac{d\theta}{4\pi}\sum_{i}(\phi_{\alpha}(\theta_{i})-\phi(2\theta_{i}))\log(1+e^{-\epsilon_{i}})\nonumber \\
 & +\frac{1}{2}\left(\log\det(\mathbb{I}-K^{-})-\log\det(\mathbb{I}-K^{+})\right)
\end{align}
where $\phi_{\alpha}=-i\partial_{\theta}\log R_{\alpha}(\theta)$
and the contribution of the measure and fluctuations can be easily
calculated in terms of finite matrices in any program say in Mathematica. 

In case of a boundary boundstate at $iu$ we add a term 
\[
\log(1+e^{-\epsilon_{u}})\quad;\qquad\epsilon_{u}=mL\cos u-\sum_{j}\phi(iu-\theta_{j})\log(1+e^{-\epsilon_{j}})
\]

\subsection{Excited states}

For simplicity we analyze a standing one-particle state in the Lee-Yang
model. This means two singularities at $\bar{\theta}_{1}=i\delta_{1}$
and $\bar{\theta}_{2}=-i\delta_{1}$. We denote $\bar{\theta}_{1}$
by $\theta_{\bar{1}}$. The idea is to extend functions and kernels
with extra components at $\bar{\theta}_{1}$: $f_{\bar{1}}=f(\bar{\theta}_{1})$
and $\phi_{\bar{1}j}=\phi(\bar{\theta}_{1}-\theta_{j})$. In particular
the TBA equations take the form 
\begin{align}
\epsilon_{i} & =mL\cosh\theta_{i}+\log\frac{S(\theta_{i}-\theta_{\bar{1}})}{S(\theta_{i}+\theta_{\bar{1}})}-\sum_{j}\phi_{ij}\log(1+e^{-\epsilon_{j}})\\
0 & =mL\cosh\theta_{\bar{1}}-\log S(2\theta_{\bar{1}})-\sum_{j}\phi_{\bar{1}j}\log(1+e^{-\epsilon_{j}})\nonumber 
\end{align}
One can solve these equations by iterations starting from $\epsilon_{i}=0$
and $\theta_{\bar{1}}=i(\pi/6+\sqrt{3}e^{-mL\cos(\pi/6)})$. Once
$\theta_{\bar{1}}$ and $\epsilon_{i}$ are obtained the energy is
\begin{equation}
E(\{0\},L)/m=2i\sinh\theta_{\bar{1}}-\frac{d\theta}{2\pi}\sum_{i}\cosh\theta_{i}\log(1+e^{-\epsilon_{i}})
\end{equation}
In calculating the excited state g-function the saddle point contribution
has the previous form, but there is an extra contribution from $\theta_{\bar{1}}$:
\begin{equation}
\log\frac{R_{\mathbb{I}}(\theta_{\bar{1}})}{\sqrt{S(2\theta_{\bar{1}})}}
\end{equation}
Finally the kernels $K^{\pm}$ have to be extended also with the discrete
mode to have $K_{\mathrm{ex}}^{\pm}$: 
\begin{equation}
K_{\bar{1}\bar{1}}^{\pm}=\frac{i}{\partial\epsilon_{\bar{1}}}\phi_{\bar{1}\bar{1}}^{\pm}\quad;\qquad K_{j\bar{1}}^{\pm}=\frac{i}{\partial\epsilon_{\bar{1}}}\phi_{j\bar{1}}^{\pm}
\end{equation}

\begin{equation}
K_{\bar{1}j}^{\pm}=\frac{d\theta}{2\pi}\frac{\phi_{\bar{1}j}^{\pm}}{1+e^{\epsilon_{j}}}\quad;\qquad K_{ij}^{\pm}=\frac{d\theta}{2\pi}\frac{\phi_{ij}^{\pm}}{1+e^{\epsilon_{j}}}
\end{equation}
Here $\partial\epsilon_{\bar{1}}=\partial_{\theta}\epsilon\vert_{\theta=\theta_{\bar{1}}}$.
In calculating this expression we take the derivative of the TBA equation
wrt. $\theta$: 
\begin{equation}
\partial_{\theta}\epsilon(\theta)=mL\sinh\theta+i(\phi(\theta-\bar{\theta}_{1})-\phi(\theta+\bar{\theta}_{1})+\int_{-\infty}^{\infty}\frac{d\theta'}{2\pi}\frac{\phi(\theta-\theta')}{1+e^{\epsilon(\theta')}}\partial_{\theta'}\epsilon(\theta')
\end{equation}
Thus in our discretized formulation 
\begin{align}
\partial\epsilon_{j} & =(\mathbb{I}-K)_{jk}^{-1}\text{(}mL\sinh\theta_{k}+i\phi_{k\bar{1}}^{-})\\
\partial\epsilon_{\bar{1}} & =mL\sinh\theta_{\bar{1}}+i\phi_{\bar{1}\bar{1}}^{-}+K_{\bar{1}j}\partial\epsilon_{j}\nonumber 
\end{align}
where $K_{ij}=\frac{d\theta}{2\pi}\frac{\phi_{ij}}{1+e^{\epsilon_{j}}}$
acts on the whole line. The contribution of the determinants is the
same as before except we replace $K^{\pm}$ with $K_{\mathrm{ex}}^{\pm}$.
Thus the final results for the excited state g-function is 
\begin{align}
\log g_{\alpha}^{\mathrm{ex}}= & -\frac{1}{4}\log(1+e^{-\epsilon_{i0}})+\frac{d\theta}{4\pi}\sum_{i}(\phi_{\alpha}(\theta_{i})-\phi(2\theta_{i}))\log(1+e^{-\epsilon_{i}})\\
 & +\frac{1}{2}\left(\log\det(\mathbb{I}-K_{\mathrm{ex}}^{-})-\log\det(\mathbb{I}-K_{\mathrm{ex}}^{+})\right)+\log\frac{R_{\mathbb{I}}(\theta_{\bar{1}})}{\sqrt{S(2\theta_{\bar{1}})}}\nonumber 
\end{align}
In case of a boundary boundstate at $iu$ we add a term 
\begin{equation}
\log(1+e^{-\epsilon_{u}})\quad;\qquad\epsilon_{u}=mL\cos u+\log\frac{S(iu-\theta_{\bar{1}})}{S(iu+\theta_{\bar{1}})}-\sum_{j}\phi(iu-\theta_{j})\log(1+e^{-\epsilon_{j}})
\end{equation}


\begin{thebibliography}{10}
\expandafter\ifx\csname url\endcsname\relax
  \def\url#1{\texttt{#1}}\fi
\expandafter\ifx\csname urlprefix\endcsname\relax\def\urlprefix{URL }\fi
\expandafter\ifx\csname href\endcsname\relax
  \def\href#1#2{#2} \def\path#1{#1}\fi

\bibitem{deLeeuw:2015hxa}
M.~de~Leeuw, C.~Kristjansen, K.~Zarembo, {One-point Functions in Defect CFT and
  Integrability}, JHEP 08 (2015) 098.
\newblock \href {http://arxiv.org/abs/1506.06958} {\path{arXiv:1506.06958}},
  \href {http://dx.doi.org/10.1007/JHEP08(2015)098}
  {\path{doi:10.1007/JHEP08(2015)098}}.

\bibitem{Buhl-Mortensen:2015gfd}
I.~Buhl-Mortensen, M.~de~Leeuw, C.~Kristjansen, K.~Zarembo, {One-point
  Functions in AdS/dCFT from Matrix Product States}, JHEP 02 (2016) 052.
\newblock \href {http://arxiv.org/abs/1512.02532} {\path{arXiv:1512.02532}},
  \href {http://dx.doi.org/10.1007/JHEP02(2016)052}
  {\path{doi:10.1007/JHEP02(2016)052}}.

\bibitem{deLeeuw:2018mkd}
M.~De~Leeuw, C.~Kristjansen, G.~Linardopoulos, {Scalar one-point functions and
  matrix product states of AdS/dCFT}, Phys. Lett. B 781 (2018) 238--243.
\newblock \href {http://arxiv.org/abs/1802.01598} {\path{arXiv:1802.01598}},
  \href {http://dx.doi.org/10.1016/j.physletb.2018.03.083}
  {\path{doi:10.1016/j.physletb.2018.03.083}}.

\bibitem{deLeeuw:2016ofj}
M.~de~Leeuw, C.~Kristjansen, G.~Linardopoulos, {One-point functions of
  non-protected operators in the SO(5) symmetric D3--D7 dCFT}, J. Phys. A
  50~(25) (2017) 254001.
\newblock \href {http://arxiv.org/abs/1612.06236} {\path{arXiv:1612.06236}},
  \href {http://dx.doi.org/10.1088/1751-8121/aa714b}
  {\path{doi:10.1088/1751-8121/aa714b}}.

\bibitem{deLeeuw:2019ebw}
M.~De~Leeuw, T.~Gombor, C.~Kristjansen, G.~Linardopoulos, B.~Pozsgay, {Spin
  Chain Overlaps and the Twisted Yangian}, JHEP 01 (2020) 176.
\newblock \href {http://arxiv.org/abs/1912.09338} {\path{arXiv:1912.09338}},
  \href {http://dx.doi.org/10.1007/JHEP01(2020)176}
  {\path{doi:10.1007/JHEP01(2020)176}}.

\bibitem{Buhl-Mortensen:2016pxs}
I.~Buhl-Mortensen, M.~de~Leeuw, A.~C. Ipsen, C.~Kristjansen, M.~Wilhelm,
  {One-loop one-point functions in gauge-gravity dualities with defects}, Phys.
  Rev. Lett. 117~(23) (2016) 231603.
\newblock \href {http://arxiv.org/abs/1606.01886} {\path{arXiv:1606.01886}},
  \href {http://dx.doi.org/10.1103/PhysRevLett.117.231603}
  {\path{doi:10.1103/PhysRevLett.117.231603}}.

\bibitem{Buhl-Mortensen:2017ind}
I.~Buhl-Mortensen, M.~de~Leeuw, A.~C. Ipsen, C.~Kristjansen, M.~Wilhelm,
  {Asymptotic One-Point Functions in Gauge-String Duality with Defects}, Phys.
  Rev. Lett. 119~(26) (2017) 261604.
\newblock \href {http://arxiv.org/abs/1704.07386} {\path{arXiv:1704.07386}},
  \href {http://dx.doi.org/10.1103/PhysRevLett.119.261604}
  {\path{doi:10.1103/PhysRevLett.119.261604}}.

\bibitem{Linardopoulos:2020jck}
G.~Linardopoulos, {Solving holographic defects}, PoS CORFU2019 (2019) 141.
\newblock \href {http://arxiv.org/abs/2005.02117} {\path{arXiv:2005.02117}}.

\bibitem{Kristjansen:2020mhn}
C.~Kristjansen, D.~M\"uller, K.~Zarembo, {Integrable boundary states in D3-D5
  dCFT: beyond scalars}, JHEP 08 (2020) 103.
\newblock \href {http://arxiv.org/abs/2005.01392} {\path{arXiv:2005.01392}},
  \href {http://dx.doi.org/10.1007/JHEP08(2020)103}
  {\path{doi:10.1007/JHEP08(2020)103}}.

\bibitem{Gombor:2020kgu}
T.~Gombor, Z.~Bajnok, {Boundary states, overlaps, nesting and bootstrapping
  AdS/dCFT}, JHEP 10 (2020) 123.
\newblock \href {http://arxiv.org/abs/2004.11329} {\path{arXiv:2004.11329}},
  \href {http://dx.doi.org/10.1007/JHEP10(2020)123}
  {\path{doi:10.1007/JHEP10(2020)123}}.

\bibitem{Gombor:2020auk}
T.~Gombor, Z.~Bajnok, {Boundary state bootstrap and asymptotic overlaps in
  AdS/dCFT}\href {http://arxiv.org/abs/2006.16151} {\path{arXiv:2006.16151}}.

\bibitem{Komatsu:2020sup}
S.~Komatsu, Y.~Wang, {Non-perturbative defect one-point functions in planar
  $\mathcal{N}=4$ super-Yang-Mills}, Nucl. Phys. B 958 (2020) 115120.
\newblock \href {http://arxiv.org/abs/2004.09514} {\path{arXiv:2004.09514}},
  \href {http://dx.doi.org/10.1016/j.nuclphysb.2020.115120}
  {\path{doi:10.1016/j.nuclphysb.2020.115120}}.

\bibitem{Jiang:2019xdz}
Y.~Jiang, S.~Komatsu, E.~Vescovi, {Structure constants in $ \mathcal{N} $ = 4
  SYM at finite coupling as worldsheet g-function}, JHEP 07~(07) (2020) 037.
\newblock \href {http://arxiv.org/abs/1906.07733} {\path{arXiv:1906.07733}},
  \href {http://dx.doi.org/10.1007/JHEP07(2020)037}
  {\path{doi:10.1007/JHEP07(2020)037}}.

\bibitem{Jiang:2019zig}
Y.~Jiang, S.~Komatsu, E.~Vescovi, {Exact Three-Point Functions of Determinant
  Operators in Planar $N=4$ Supersymmetric Yang-Mills Theory}, Phys. Rev. Lett.
  123~(19) (2019) 191601.
\newblock \href {http://arxiv.org/abs/1907.11242} {\path{arXiv:1907.11242}},
  \href {http://dx.doi.org/10.1103/PhysRevLett.123.191601}
  {\path{doi:10.1103/PhysRevLett.123.191601}}.

\bibitem{Caux:2013ra}
J.-S. Caux, F.~H. Essler, {Time evolution of local observables after quenching
  to an integrable model}, Phys. Rev. Lett. 110~(25) (2013) 257203.
\newblock \href {http://arxiv.org/abs/1301.3806} {\path{arXiv:1301.3806}},
  \href {http://dx.doi.org/10.1103/PhysRevLett.110.257203}
  {\path{doi:10.1103/PhysRevLett.110.257203}}.

\bibitem{Kozlowski:2012fv}
K.~Kozlowski, B.~Pozsgay, {Surface free energy of the open XXZ spin-1/2 chain},
  J. Stat. Mech. 1205 (2012) P05021.
\newblock \href {http://arxiv.org/abs/1201.5884} {\path{arXiv:1201.5884}},
  \href {http://dx.doi.org/10.1088/1742-5468/2012/05/P05021}
  {\path{doi:10.1088/1742-5468/2012/05/P05021}}.

\bibitem{Brockmann_2014_1}
M.~Brockmann, J.~De~Nardis, B.~Wouters, J.-S. Caux, {A Gaudin-like determinant
  for overlaps of Neel and XXZ Bethe states}, Journal of Physics A:
  Mathematical and Theoretical 47~(14) (2014) 145003.
\newblock \href {http://dx.doi.org/10.1088/1751-8113/47/14/145003}
  {\path{doi:10.1088/1751-8113/47/14/145003}}.

\bibitem{Pozsgay:2018ixm}
B.~Pozsgay, {Overlaps with arbitrary two-site states in the XXZ spin chain}, J.
  Stat. Mech. 1805~(5) (2018) 053103.
\newblock \href {http://arxiv.org/abs/1801.03838} {\path{arXiv:1801.03838}},
  \href {http://dx.doi.org/10.1088/1742-5468/aabbe1}
  {\path{doi:10.1088/1742-5468/aabbe1}}.

\bibitem{Piroli:2017sei}
L.~Piroli, B.~Pozsgay, E.~Vernier, {What is an integrable quench?}, Nucl. Phys.
  B925 (2017) 362--402.
\newblock \href {http://arxiv.org/abs/1709.04796} {\path{arXiv:1709.04796}},
  \href {http://dx.doi.org/10.1016/j.nuclphysb.2017.10.012}
  {\path{doi:10.1016/j.nuclphysb.2017.10.012}}.

\bibitem{Piroli:2018ksf}
L.~Piroli, E.~Vernier, P.~Calabrese, B.~Pozsgay, {Integrable quenches in nested
  spin chains I: the exact steady states}, J. Stat. Mech. 1906~(6) (2019)
  063103.
\newblock \href {http://arxiv.org/abs/1811.00432} {\path{arXiv:1811.00432}},
  \href {http://dx.doi.org/10.1088/1742-5468/ab1c51}
  {\path{doi:10.1088/1742-5468/ab1c51}}.

\bibitem{Piroli:2018don}
L.~Piroli, E.~Vernier, P.~Calabrese, B.~Pozsgay, {Integrable quenches in nested
  spin chains II: fusion of boundary transfer matrices}, J. Stat. Mech.
  1906~(6) (2019) 063104.
\newblock \href {http://arxiv.org/abs/1812.05330} {\path{arXiv:1812.05330}},
  \href {http://dx.doi.org/10.1088/1742-5468/ab1c52}
  {\path{doi:10.1088/1742-5468/ab1c52}}.

\bibitem{Horvath:2015rya}
D.~Horvath, S.~Sotiriadis, G.~Takacs, {Initial states in integrable quantum
  field theory quenches from an integral equation hierarchy}, Nucl. Phys. B 902
  (2016) 508--547.
\newblock \href {http://arxiv.org/abs/1510.01735} {\path{arXiv:1510.01735}},
  \href {http://dx.doi.org/10.1016/j.nuclphysb.2015.11.025}
  {\path{doi:10.1016/j.nuclphysb.2015.11.025}}.

\bibitem{Horvath:2017wzf}
D.~Horvath, G.~Takacs, {Overlaps after quantum quenches in the sine-Gordon
  model}, Phys. Lett. B 771 (2017) 539--545.
\newblock \href {http://arxiv.org/abs/1704.00594} {\path{arXiv:1704.00594}},
  \href {http://dx.doi.org/10.1016/j.physletb.2017.05.087}
  {\path{doi:10.1016/j.physletb.2017.05.087}}.

\bibitem{Horvath:2018gat}
D.~Horvath, M.~Kormos, G.~Takacs, {Overlap singularity and time evolution in
  integrable quantum field theory}, JHEP 08 (2018) 170.
\newblock \href {http://arxiv.org/abs/1805.08132} {\path{arXiv:1805.08132}},
  \href {http://dx.doi.org/10.1007/JHEP08(2018)170}
  {\path{doi:10.1007/JHEP08(2018)170}}.

\bibitem{Rakovszky:2016ugs}
T.~Rakovszky, M.~Mestyan, M.~Collura, M.~Kormos, G.~Takacs, {Hamiltonian
  truncation approach to quenches in the Ising field theory}, Nucl. Phys. B 911
  (2016) 805--845.
\newblock \href {http://arxiv.org/abs/1607.01068} {\path{arXiv:1607.01068}},
  \href {http://dx.doi.org/10.1016/j.nuclphysb.2016.08.024}
  {\path{doi:10.1016/j.nuclphysb.2016.08.024}}.

\bibitem{Hodsagi:2019rcs}
K.~Hodsagi, M.~Kormos, G.~Takacs, {Perturbative post-quench overlaps in Quantum
  Field Theory}, JHEP 08 (2019) 047.
\newblock \href {http://arxiv.org/abs/1905.05623} {\path{arXiv:1905.05623}},
  \href {http://dx.doi.org/10.1007/JHEP08(2019)047}
  {\path{doi:10.1007/JHEP08(2019)047}}.

\bibitem{Kormos:2010ae}
M.~Kormos, B.~Pozsgay, {One-Point Functions in Massive Integrable QFT with
  Boundaries}, JHEP 04 (2010) 112.
\newblock \href {http://arxiv.org/abs/1002.2783} {\path{arXiv:1002.2783}},
  \href {http://dx.doi.org/10.1007/JHEP04(2010)112}
  {\path{doi:10.1007/JHEP04(2010)112}}.

\bibitem{Takacs:2011aa}
G.~Takacs, G.~Watts, {Excited State G-Functions from the Truncated Conformal
  Space}, JHEP 02 (2012) 082.
\newblock \href {http://arxiv.org/abs/1112.2906} {\path{arXiv:1112.2906}},
  \href {http://dx.doi.org/10.1007/JHEP02(2012)082}
  {\path{doi:10.1007/JHEP02(2012)082}}.

\bibitem{Dorey:1997yg}
P.~Dorey, A.~Pocklington, R.~Tateo, G.~Watts, {TBA and TCSA with boundaries and
  excited states}, Nucl. Phys. B525 (1998) 641--663.
\newblock \href {http://arxiv.org/abs/hep-th/9712197}
  {\path{arXiv:hep-th/9712197}}, \href
  {http://dx.doi.org/10.1016/S0550-3213(98)00339-3}
  {\path{doi:10.1016/S0550-3213(98)00339-3}}.

\bibitem{Dorey:1999cj}
P.~Dorey, I.~Runkel, R.~Tateo, G.~Watts, {g function flow in perturbed boundary
  conformal field theories}, Nucl. Phys. B 578 (2000) 85--122.
\newblock \href {http://arxiv.org/abs/hep-th/9909216}
  {\path{arXiv:hep-th/9909216}}, \href
  {http://dx.doi.org/10.1016/S0550-3213(99)00772-5}
  {\path{doi:10.1016/S0550-3213(99)00772-5}}.

\bibitem{Luscher:1985dn}
M.~Luscher, {Volume Dependence of the Energy Spectrum in Massive Quantum Field
  Theories. 1. Stable Particle States}, Commun. Math. Phys. 104 (1986) 177.
\newblock \href {http://dx.doi.org/10.1007/BF01211589}
  {\path{doi:10.1007/BF01211589}}.

\bibitem{Luscher:1986pf}
M.~Luscher, {Volume Dependence of the Energy Spectrum in Massive Quantum Field
  Theories. 2. Scattering States}, Commun. Math. Phys. 105 (1986) 153--188.
\newblock \href {http://dx.doi.org/10.1007/BF01211097}
  {\path{doi:10.1007/BF01211097}}.

\bibitem{Bajnok:2008bm}
Z.~Bajnok, R.~A. Janik, {Four-loop perturbative Konishi from strings and finite
  size effects for multiparticle states}, Nucl. Phys. B 807 (2009) 625--650.
\newblock \href {http://arxiv.org/abs/0807.0399} {\path{arXiv:0807.0399}},
  \href {http://dx.doi.org/10.1016/j.nuclphysb.2008.08.020}
  {\path{doi:10.1016/j.nuclphysb.2008.08.020}}.

\bibitem{Zamolodchikov:1989cf}
A.~Zamolodchikov, {Thermodynamic Bethe Ansatz in Relativistic Models. Scaling
  Three State Potts and Lee-yang Models}, Nucl. Phys. B 342 (1990) 695--720.
\newblock \href {http://dx.doi.org/10.1016/0550-3213(90)90333-9}
  {\path{doi:10.1016/0550-3213(90)90333-9}}.

\bibitem{Dorey:1996re}
P.~Dorey, R.~Tateo, {Excited states by analytic continuation of TBA equations},
  Nucl. Phys. B 482 (1996) 639--659.
\newblock \href {http://arxiv.org/abs/hep-th/9607167}
  {\path{arXiv:hep-th/9607167}}, \href
  {http://dx.doi.org/10.1016/S0550-3213(96)00516-0}
  {\path{doi:10.1016/S0550-3213(96)00516-0}}.

\bibitem{Affleck:1991tk}
I.~Affleck, A.~W. Ludwig, {Universal noninteger 'ground state degeneracy' in
  critical quantum systems}, Phys. Rev. Lett. 67 (1991) 161--164.
\newblock \href {http://dx.doi.org/10.1103/PhysRevLett.67.161}
  {\path{doi:10.1103/PhysRevLett.67.161}}.

\bibitem{Friedan:2003yc}
D.~Friedan, A.~Konechny, {On the boundary entropy of one-dimensional quantum
  systems at low temperature}, Phys. Rev. Lett. 93 (2004) 030402.
\newblock \href {http://arxiv.org/abs/hep-th/0312197}
  {\path{arXiv:hep-th/0312197}}, \href
  {http://dx.doi.org/10.1103/PhysRevLett.93.030402}
  {\path{doi:10.1103/PhysRevLett.93.030402}}.

\bibitem{LeClair:1995uf}
A.~LeClair, G.~Mussardo, H.~Saleur, S.~Skorik, {Boundary energy and boundary
  states in integrable quantum field theories}, Nucl. Phys. B 453 (1995)
  581--618.
\newblock \href {http://arxiv.org/abs/hep-th/9503227}
  {\path{arXiv:hep-th/9503227}}, \href
  {http://dx.doi.org/10.1016/0550-3213(95)00435-U}
  {\path{doi:10.1016/0550-3213(95)00435-U}}.

\bibitem{Woynarovich:2004gc}
F.~Woynarovich, {O(1) contribution of saddle point fluctuations to the free
  energy of Bethe Ansatz systems}, Nucl. Phys. B 700 (2004) 331--360.
\newblock \href {http://arxiv.org/abs/cond-mat/0402129}
  {\path{arXiv:cond-mat/0402129}}, \href
  {http://dx.doi.org/10.1016/j.nuclphysb.2004.08.043}
  {\path{doi:10.1016/j.nuclphysb.2004.08.043}}.

\bibitem{Dorey:2004xk}
P.~Dorey, D.~Fioravanti, C.~Rim, R.~Tateo, {Integrable quantum field theory
  with boundaries: The Exact g function}, Nucl. Phys. B 696 (2004) 445--467.
\newblock \href {http://arxiv.org/abs/hep-th/0404014}
  {\path{arXiv:hep-th/0404014}}, \href
  {http://dx.doi.org/10.1016/j.nuclphysb.2004.06.045}
  {\path{doi:10.1016/j.nuclphysb.2004.06.045}}.

\bibitem{Pozsgay:2010tv}
B.~Pozsgay, {On O(1) contributions to the free energy in Bethe Ansatz systems:
  The Exact g-function}, JHEP 08 (2010) 090.
\newblock \href {http://arxiv.org/abs/1003.5542} {\path{arXiv:1003.5542}},
  \href {http://dx.doi.org/10.1007/JHEP08(2010)090}
  {\path{doi:10.1007/JHEP08(2010)090}}.

\bibitem{Kostov:2018dmi}
I.~Kostov, D.~Serban, D.-L. Vu, {Boundary TBA, trees and loops}, Nucl. Phys. B
  949 (2019) 114817.
\newblock \href {http://arxiv.org/abs/1809.05705} {\path{arXiv:1809.05705}},
  \href {http://dx.doi.org/10.1016/j.nuclphysb.2019.114817}
  {\path{doi:10.1016/j.nuclphysb.2019.114817}}.

\bibitem{Kostov:2019sgu}
I.~Kostov, {Effective Quantum Field Theory for the Thermodynamical Bethe
  Ansatz}, JHEP 02 (2020) 043.
\newblock \href {http://arxiv.org/abs/1911.07343} {\path{arXiv:1911.07343}},
  \href {http://dx.doi.org/10.1007/JHEP02(2020)043}
  {\path{doi:10.1007/JHEP02(2020)043}}.

\bibitem{Ghoshal:1993tm}
S.~Ghoshal, A.~B. Zamolodchikov, {Boundary S matrix and boundary state in
  two-dimensional integrable quantum field theory}, Int. J. Mod. Phys. A 9
  (1994) 3841--3886, [Erratum: Int.J.Mod.Phys.A 9, 4353 (1994)].
\newblock \href {http://arxiv.org/abs/hep-th/9306002}
  {\path{arXiv:hep-th/9306002}}, \href
  {http://dx.doi.org/10.1142/S0217751X94001552}
  {\path{doi:10.1142/S0217751X94001552}}.

\bibitem{Cardy:1989fw}
J.~L. Cardy, G.~Mussardo, {S Matrix of the Yang-Lee Edge Singularity in
  Two-Dimensions}, Phys. Lett. B225 (1989) 275--278.
\newblock \href {http://dx.doi.org/10.1016/0370-2693(89)90818-6}
  {\path{doi:10.1016/0370-2693(89)90818-6}}.

\bibitem{Dorey:1998kt}
P.~Dorey, R.~Tateo, G.~Watts, {Generalizations of the Coleman-Thun mechanism
  and boundary reflection factors}, Phys. Lett. B 448 (1999) 249--256.
\newblock \href {http://arxiv.org/abs/hep-th/9810098}
  {\path{arXiv:hep-th/9810098}}, \href
  {http://dx.doi.org/10.1016/S0370-2693(99)00004-0}
  {\path{doi:10.1016/S0370-2693(99)00004-0}}.

\bibitem{Bajnok:2006dn}
Z.~Bajnok, L.~Palla, G.~Takacs, {Boundary one-point function, Casimir energy
  and boundary state formalism in D+1 dimensional QFT}, Nucl. Phys. B 772
  (2007) 290--322.
\newblock \href {http://arxiv.org/abs/hep-th/0611176}
  {\path{arXiv:hep-th/0611176}}, \href
  {http://dx.doi.org/10.1016/j.nuclphysb.2007.02.023}
  {\path{doi:10.1016/j.nuclphysb.2007.02.023}}.

\bibitem{Bajnok:2004tq}
Z.~Bajnok, L.~Palla, G.~Takacs, {Finite size effects in quantum field theories
  with boundary from scattering data}, Nucl. Phys. B 716 (2005) 519--542.
\newblock \href {http://arxiv.org/abs/hep-th/0412192}
  {\path{arXiv:hep-th/0412192}}, \href
  {http://dx.doi.org/10.1016/j.nuclphysb.2005.03.021}
  {\path{doi:10.1016/j.nuclphysb.2005.03.021}}.

\bibitem{Bajnok:2006ze}
Z.~Bajnok, L.~Palla, G.~Takacs, {On the boundary form-factor program}, Nucl.
  Phys. B 750 (2006) 179--212.
\newblock \href {http://arxiv.org/abs/hep-th/0603171}
  {\path{arXiv:hep-th/0603171}}, \href
  {http://dx.doi.org/10.1016/j.nuclphysb.2006.05.019}
  {\path{doi:10.1016/j.nuclphysb.2006.05.019}}.

\bibitem{Pozsgay:2007gx}
B.~Pozsgay, G.~Takacs, {Form factors in finite volume. II. Disconnected terms
  and finite temperature correlators}, Nucl. Phys. B 788 (2008) 209--251.
\newblock \href {http://arxiv.org/abs/0706.3605} {\path{arXiv:0706.3605}},
  \href {http://dx.doi.org/10.1016/j.nuclphysb.2007.07.008}
  {\path{doi:10.1016/j.nuclphysb.2007.07.008}}.

\bibitem{Pozsgay:2007kn}
B.~Pozsgay, G.~Takacs, {Form-factors in finite volume I: Form-factor bootstrap
  and truncated conformal space}, Nucl. Phys. B 788 (2008) 167--208.
\newblock \href {http://arxiv.org/abs/0706.1445} {\path{arXiv:0706.1445}},
  \href {http://dx.doi.org/10.1016/j.nuclphysb.2007.06.027}
  {\path{doi:10.1016/j.nuclphysb.2007.06.027}}.

\bibitem{Jiang:2020sdw}
Y.~Jiang, B.~Pozsgay, {On exact overlaps in integrable spin chains}, JHEP 06
  (2020) 022.
\newblock \href {http://arxiv.org/abs/2002.12065} {\path{arXiv:2002.12065}},
  \href {http://dx.doi.org/10.1007/JHEP06(2020)022}
  {\path{doi:10.1007/JHEP06(2020)022}}.

\bibitem{Woynarovich:2010wt}
F.~Woynarovich, {On the normalization of the partition function of Bethe Ansatz
  systems}, Nucl. Phys. B 852 (2011) 269--286.
\newblock \href {http://arxiv.org/abs/1007.1148} {\path{arXiv:1007.1148}},
  \href {http://dx.doi.org/10.1016/j.nuclphysb.2011.06.015}
  {\path{doi:10.1016/j.nuclphysb.2011.06.015}}.

\bibitem{Bajnok:2019yik}
Z.~Bajnok, F.~Smirnov, {Diagonal finite volume matrix elements in the
  sinh-Gordon model}, Nucl. Phys. B 945 (2019) 114664.
\newblock \href {http://arxiv.org/abs/1903.06990} {\path{arXiv:1903.06990}},
  \href {http://dx.doi.org/10.1016/j.nuclphysb.2019.114664}
  {\path{doi:10.1016/j.nuclphysb.2019.114664}}.

\bibitem{Bajnok:2019mpp}
Z.~Bajnok, I.~Vona, {Exact finite volume expectation values of conserved
  currents}, Phys. Lett. B 805 (2020) 135446.
\newblock \href {http://arxiv.org/abs/1911.08525} {\path{arXiv:1911.08525}},
  \href {http://dx.doi.org/10.1016/j.physletb.2020.135446}
  {\path{doi:10.1016/j.physletb.2020.135446}}.

\bibitem{Yurov:1989yu}
V.~Yurov, A.~Zamolodchikov, {Truncated Conformal Space Approach to Scaling
  Lee-Yang model}, Int. J. Mod. Phys. A 5 (1990) 3221--3246.
\newblock \href {http://dx.doi.org/10.1142/S0217751X9000218X}
  {\path{doi:10.1142/S0217751X9000218X}}.

\bibitem{Kormos:2007qx}
M.~Kormos, G.~Takacs, {Boundary form-factors in finite volume}, Nucl. Phys. B
  803 (2008) 277--298.
\newblock \href {http://arxiv.org/abs/0712.1886} {\path{arXiv:0712.1886}},
  \href {http://dx.doi.org/10.1016/j.nuclphysb.2008.05.003}
  {\path{doi:10.1016/j.nuclphysb.2008.05.003}}.

\bibitem{Lencses:2011ab}
M.~Lencses, G.~Takacs, {Breather boundary form factors in sine-Gordon theory},
  Nucl. Phys. B 852 (2011) 615--633.
\newblock \href {http://arxiv.org/abs/1106.1902} {\path{arXiv:1106.1902}},
  \href {http://dx.doi.org/10.1016/j.nuclphysb.2011.07.010}
  {\path{doi:10.1016/j.nuclphysb.2011.07.010}}.

\bibitem{Cardy:1989ir}
J.~L. Cardy, {Boundary Conditions, Fusion Rules and the Verlinde Formula},
  Nucl. Phys. B324 (1989) 581--596.
\newblock \href {http://dx.doi.org/10.1016/0550-3213(89)90521-X}
  {\path{doi:10.1016/0550-3213(89)90521-X}}.

\bibitem{Konechny:2016eek}
A.~Konechny, {RG boundaries and interfaces in Ising field theory}, J. Phys. A
  50~(14) (2017) 145403.
\newblock \href {http://arxiv.org/abs/1610.07489} {\path{arXiv:1610.07489}},
  \href {http://dx.doi.org/10.1088/1751-8121/aa60f6}
  {\path{doi:10.1088/1751-8121/aa60f6}}.

\bibitem{Bajnok:2019cdf}
Z.~Bajnok, M.~Lajer, B.~Szepfalvi, I.~Vona, {Leading exponential finite size
  corrections for non-diagonal form factors}, JHEP 07 (2019) 173.
\newblock \href {http://arxiv.org/abs/1904.00492} {\path{arXiv:1904.00492}},
  \href {http://dx.doi.org/10.1007/JHEP07(2019)173}
  {\path{doi:10.1007/JHEP07(2019)173}}.

\bibitem{Konik:2020gdi}
R.~Konik, M.~L\'ajer, G.~Mussardo, {Approaching the self-dual point of the
  sinh-Gordon model}, JHEP 01 (2021) 014.
\newblock \href {http://arxiv.org/abs/2007.00154} {\path{arXiv:2007.00154}},
  \href {http://dx.doi.org/10.1007/JHEP01(2021)014}
  {\path{doi:10.1007/JHEP01(2021)014}}.

\bibitem{Caetano:2020dyp}
J.~a. Caetano, S.~Komatsu, {Functional equations and separation of variables
  for exact $g$-function}, JHEP 09 (2020) 180.
\newblock \href {http://arxiv.org/abs/2004.05071} {\path{arXiv:2004.05071}},
  \href {http://dx.doi.org/10.1007/JHEP09(2020)180}
  {\path{doi:10.1007/JHEP09(2020)180}}.

\bibitem{Kostov:2019fvw}
D.-L. Vu, I.~Kostov, D.~Serban, {Boundary entropy of integrable perturbed SU
  (2)$_{k}$ WZNW}, JHEP 08 (2019) 154.
\newblock \href {http://arxiv.org/abs/1906.01909} {\path{arXiv:1906.01909}},
  \href {http://dx.doi.org/10.1007/JHEP08(2019)154}
  {\path{doi:10.1007/JHEP08(2019)154}}.

\bibitem{Gombor:2017qsy}
T.~Gombor, {Nonstandard Bethe Ansatz equations for open O(N) spin chains},
  Nucl. Phys. B 935 (2018) 310--343.
\newblock \href {http://arxiv.org/abs/1712.03753} {\path{arXiv:1712.03753}},
  \href {http://dx.doi.org/10.1016/j.nuclphysb.2018.08.014}
  {\path{doi:10.1016/j.nuclphysb.2018.08.014}}.

\bibitem{Bajnok:2020xoz}
Z.~Bajnok, J.~L. Jacobsen, Y.~Jiang, R.~I. Nepomechie, Y.~Zhang, {Cylinder
  partition function of the 6-vertex model from algebraic geometry}, JHEP 06
  (2020) 169.
\newblock \href {http://arxiv.org/abs/2002.09019} {\path{arXiv:2002.09019}},
  \href {http://dx.doi.org/10.1007/JHEP06(2020)169}
  {\path{doi:10.1007/JHEP06(2020)169}}.

\end{thebibliography}
\end{document}